\documentclass[%
 aip,
 amsmath,amssymb,
 reprint,longbibliography,
]{revtex4-1}
\usepackage[utf8]{inputenc}
\usepackage{graphicx}
\usepackage{bbold}
\usepackage{cases}
\usepackage{upgreek}
\usepackage{bm}
\usepackage[usenames, dvipsnames]{color}
\usepackage[colorlinks=true, citecolor=Thistle, linkcolor=JungleGreen, urlcolor=Sepia]{hyperref}
\usepackage{braket}
\usepackage[normalem]{ulem}
\usepackage{xfrac}
\usepackage{subcaption} 
\usepackage{ragged2e} 
\usepackage{spverbatim}  
\usepackage{rotating}  
\DeclareCaptionJustification{justified}{\justifying}

\usepackage{textgreek}
\usepackage{gensymb}
\usepackage{dcolumn}

\newcommand \mc[1] { \mathcal{#1} }
\newcommand \dd[1]  { \!\textrm d{#1} \,}   
\newcommand \rmm[1]  { \textrm{#1} }

\newcommand \e[1] { \rmm{e}^{#1} }

\newcommand \lr[1] { \left \langle #1 \right \rangle }

\newcommand \imi { \mathrm{i}}

\makeatletter
\def\@email#1#2{%
 \endgroup
 \patchcmd{\titleblock@produce}
  {\frontmatter@RRAPformat}
  {\frontmatter@RRAPformat{\produce@RRAP{*#1\href{mailto:#2}{#2}}}\frontmatter@RRAPformat}
  {}{}
}%
\makeatother
\begin{document}

\preprint{AIP/123-QED}

\title{Data-driven, non-Markovian modelling of weather in the presence of non-stationary, non-Gaussian, and heteroskedastic climate dynamics}

\author{Thomas Sayer}
\homepage{thomas.e.sayer@durham.ac.uk}
\affiliation{Department of Chemistry, Durham University, Durham, DH1 3LE, United Kingdom\looseness=-1} 

\author{Andr\'{e}s Montoya-Castillo}
\homepage{Andres.MontoyaCastillo@colorado.edu}
\affiliation{Department of Chemistry, University of Colorado Boulder, Boulder, CO 80309, USA\looseness=-1} 

\date{\today}

\begin{abstract}
While the generalized Langevin equation (GLE) is a powerful tool to understand the behavior of complex dissipative systems, driving by external fields renders standard GLE construction workflows invalid. Filtering approaches that separate fluctuations from the non-equilibrium response can sometimes circumvent the need for a non-equilibrium formalism when the residual fluctuations are homoskedastic, stationary, and preferably Gaussian. Here, we introduce the temperature time series from Boulder, Colorado, as representative of the more general and complex case where the filtered time series remains non-Gaussian, non-stationary, and heteroskedastic. With this example, we develop a protocol to build an accurate and efficient low-dimensional description of the weather fluctuations. Our protocol classifies the weather data based on the position in the annual cycle, and introduces local homoskedasticity as a metric to identify seasons of likely stationarity. Within these seasons, we build pseudo-equilibrium models. Leveraging state-based generalized master equation modelling as an alternative to the GLE, we resolve difficulties like non-Gaussianity and position dependence of the memory (friction) kernel. Our data-driven approach accurately reproduces the evolving fluctuations of the Boulder temperature time series, illustrating the feasibility of our method as a general tool to describe driven, dissipative systems.
\end{abstract}

\maketitle

\section{Introduction}
Simulating the dynamics of driven, dissipative many-body systems remains a fundamental theoretical and computational challenge. These systems are ubiquitous, and the ability to accurately and efficiently predict their dynamics is essential across meteorology, seismology, financial markets, ecology, chemistry, and physics. What makes predicting the dynamics of these systems difficult is the complex emergent behavior that develops across multiple scales. A powerful and appealing option is to invoke dimensionality reduction tools that provide a principled low-dimensional equation describing the evolution of a small set of variables of interest to experiment or theory. Arguably most famous among these tools is the generalized Langevin equation (GLE),\cite{BookZwanzig} often written in its `approximate' form as
\begin{equation}\label{eq:GLE_approx}
    \ddot{x}_t = -\frac{\nabla_x \Phi_\mathrm{MF}(x_t)}{M} - \int_0^t\dd{s}\dot{x}_s\Gamma_\mathrm{app}(t-s) + f_\mathrm{Z}^\mathrm{R}(t) ,   
\end{equation}
which offers a stochastic differential equation of motion by adding to the mean force, $-\nabla_x \Phi_\mathrm{MF}(x_t)$, a random force, $f_\mathrm{Z}^\mathrm{R}(t)$, that encodes instantaneous interactions from all other particles in the many-body system, and an associated time-dependent friction kernel, $\Gamma_\mathrm{app}(t-s)$, that dissipates energy. As the ``approximate'' label suggests, GLEs are notoriously difficult to construct exactly. 

Why is it hard to construct a GLE? Practically, converging the needed quantities is statistically expensive, even for systems at equilibrium. While various algorithms to extract the kernel and random force can be adapted to suit the particular problem,\cite{Harp1970, BookBoonYip, Carof2014b, Jung2017, Daldrop2018} this challenge becomes foundational in systems with limited observations, like weather and climate. To address this challenge, data-driven approaches have become popular,\cite{Chorin2015, Lei2016, Vroylandt2022a, Lyu2023, Xie2024} but even these struggle when the data are sparse and noisy. Separately, there are subtle difficulties in the mathematical derivation, even at equilibrium. For example, the approximate GLE (Eq.~\ref{eq:GLE_approx}) neglects $x_t$-dependence in the mass, and both $x_t$- and $\dot{x}_t$-dependence in the friction (see Appendix~\ref{app:GLE}).\cite{Glatzel2021, Ayaz2022, Vroylandt2022b} Furthermore, the random force is often assumed to be a Gaussian process, even when this may not be appropriate.\cite{Ayaz2022, Netz2023} Although these simplifications can often be justified in model Hamiltonians, there is no guarantee that they remain so in real life. 

The non-equilibrium problem represents a further, active frontier for the GLE.\cite{Cui2018, TeVrugt2019, Meyer2019, Schilling2022, Netz2024, Hery2024, Koch2024} Out of equilibrium, one must contend with the loss of detailed balance and fluctuation-dissipation theorems that would otherwise fix the relation between noise and friction. Hence, because the standard workflow to utilize Eq.~\ref{eq:GLE_approx} relies on translational invariance of the fluctuating time series, applying an equilibrium workflow to extract the memory kernel and random force to a driven system yields objects contaminated by the forcing. Careful dissection of the different contributions then becomes necessary. Yet, the equilibrium toolbox might still be useable if the steady-state component could be removed to reveal fluctuations with pseudo-equilibrium statistics.

Protocols have emerged to filter weather and climate data to permit a simple mathematical description of the residual dynamics.\cite{Galanis2006, Kwaniok2009, Kwasniok2013, Hongli2019, Netz2024, Hery2024} In some cases, these can remove the external driving, yielding a time series that appears to be produced by a stationary process, and render it amenable to a GLE description. For example, Fourier filtering\cite{Netz2025} of the temperature time series at Berlin-Tegel airport between January 15, 2006, and February 22, 2010, gave a pseudo-equilibrium trajectory that could be analyzed with an \textit{equilibrium} Mori-type GLE.\cite{Kiefer2025} That is, while the system dictating surface temperature dynamics constitutes a driven, dissipative many-body system,\cite{Giorgini2025, Falasca2025, Falasca2024, Ghil2020} filtering the annual parts of the time series yielded temperature fluctuations that were time translationally invariant and even normally distributed,
\footnote{At least over the timescales tested. There is a slight slope to the background from global warming, and it is known that longer timescale processes are also at play.\cite{Ghil2020, Todd2025}}
suggesting an underlying stationary Gaussian process. The consequences were powerful: one could construct a GLE to describe the temporal evolution of temperature fluctuations that is better than standard machine learning tools.\cite{Kiefer2025} But how general is this result?

To address this question, we must understand when and how low-dimensional dynamics projected from driven many-body systems may be reliably simplified. Here, the Floquet theorem ensures that for periodically driven systems that are in the Markovian limit before driving, factorization of the steady state recovers residual dynamics that are again Markovian.\cite{Horenko2001, Wang2015, Knoch2019} More generally, the simplification of stationarity arises when a driven system reaches a nonequilibrium steady state after some transient response time.\cite{Komatsu2009, Chou2011, Seifert2012} One expects this to happen if the driving frequency is sufficiently timescale-separated from that of the fluctuations. This holds even though the fast dynamics may be different from those that would exist for the system in the absence of forcing. The temperature dynamics at the Berlin-Tegel airport lie in this regime. However, timescale separation in weather prediction and climate science is rarely possible.\cite{Santer2011, Palmer2019, Ghil2020, Lucarini2023} In the absence of this simplification, the fluctuation-dissipation component will depend parametrically on the slow drive. In other words, we anticipate the absence of a separable non-equilibrium steady state, and the emergence of time-dependent fluctuation statistics, or heteroskedasticity.


\begin{figure*}[!ht]
    \vspace{-12pt}
    \centering
    \begin{subfigure}[b]{0.4\textwidth}
        \resizebox{\textwidth}{!}{\includegraphics{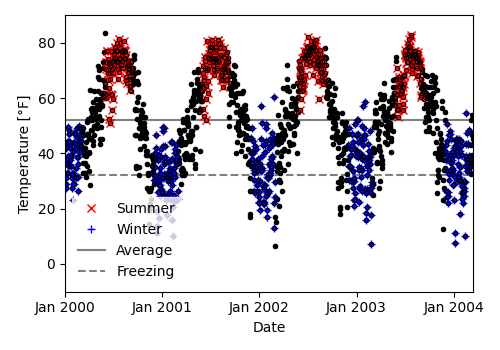}} 
        \caption{}
        \label{fig:1a}
    \end{subfigure}
    \begin{subfigure}[b]{0.4\textwidth}
        \resizebox{\textwidth}{!}{\includegraphics{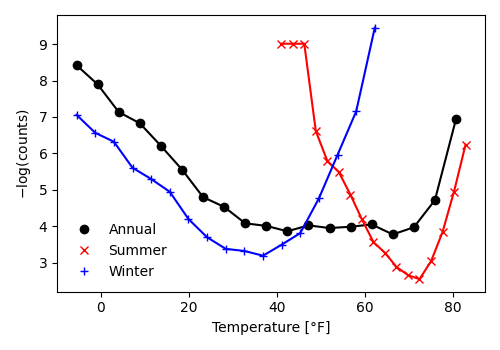}} 
        \caption{}
        \label{fig:1b}
    \end{subfigure}
    \\
    \begin{subfigure}[b]{0.4\textwidth}
        \resizebox{\textwidth}{!}{\includegraphics{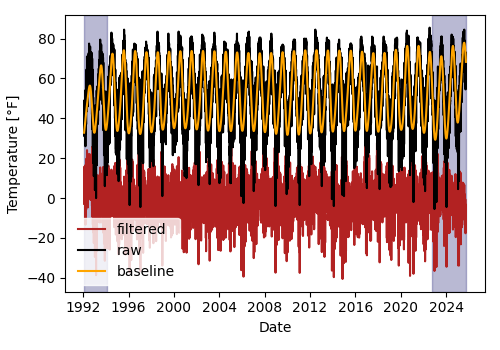}} 
        \caption{}
        \label{fig:1c}
    \end{subfigure}
    \begin{subfigure}[b]{0.4\textwidth}
        \resizebox{\textwidth}{!}{\includegraphics{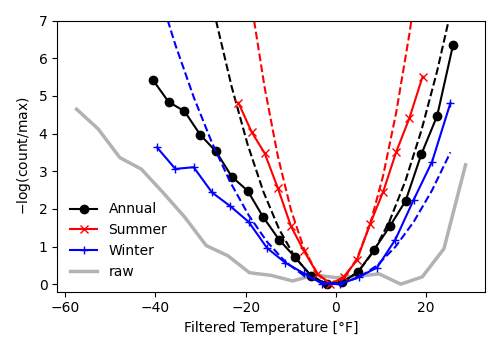}} 
        \caption{}
        \label{fig:1d}
    \end{subfigure}
    
    \caption{Daily average temperature data from Boulder, Colorado, from 1992 to 2025.\cite{NOAA_Boulder} \textbf{(a)} Four-year window of the raw data, with the traditional summer and winter calendar months highlighted with red and blue markers, respectively. The spread in the winter is much larger. \textbf{(b)} Histogram of the raw data showing a bistable \textit{and strongly asymmetric} distribution. Histograms of just the summer or winter months have different forms. \textbf{(c)} Filtered time series data showing the fluctuations in red and the residual periodic ``baseline'' in orange. Grey bars show regions excluded from our analysis due to filter-induced artifacts. See Fig.~\ref{fig:correlations} for further details. \textbf{(d)} Filtered fluctuations become unimodal but remain asymmetric. Again, the statistics differ qualitatively when considering only summer or winter months.}
    \label{fig:dataset}
\end{figure*}

\section{Filtering and the equilibrium GLE}\label{sec:filtering}
The success of the Fourier-filtered GLE\cite{Kiefer2025} is striking. It paints a picture of temperature dynamics dominated by deterministic annual driving with unassuming Gaussian fluctuations on top. This means that at any time of year, a warmer day is as likely as a colder day---an implication that extends to extreme weather events. To test the robustness of this result, we processed the temperature time series data for two other locations. First, employing the same workflow,\cite{Kiefer2025} we examined data from Durham, England,\cite{Burt2022} and found the same symmetric, Gaussian result. As Berlin and Durham both have North European weather, this similarity may not be too surprising. Yet, Gaussian statistics have also been shown to hold for the detrended average \textit{global} surface temperature.\cite{Lak2025} So, does the result still hold for the local temperature somewhere with an unusual geography? Say, Boulder, Colorado, at the foothills of the Rocky Mountains, with a high altitude ($\sim 1$~mile above sea level) and around more central latitude (40 degrees north) around 15 degrees further south than Berlin and Durham. Boulder is known to have somewhat variable weather.

To perform the reduced dimensionality analysis, we extract several different objects which we introduce below (see Appendix~\ref{app:methods} for details). This approach requires that we assume the \textit{equilibrium} GLE applies to the filtered data. Commonly, the GLE observable $\dot{X}(t)$ is the velocity of some particle, selected along with its position via the projector, $\mc{P}$. We select the temperature at our weather station as $x_t\equiv T(t)$. The general form reads
\begin{equation}\label{eq:GLE}
    \ddot{T}(t)
    =
    \underbrace{ \mathrm{i} \mc{PL} \dot{T}(t) }_{\text{effective force}}
    +
    \underbrace{ \int_0^t \dd{s} \dot{T}(s)\mc{K}(t-s) }_{\text{memory (friction)}}
    +
    \underbrace{ \e{\mathrm{i}\mc{QL}t} \mc{QL} \dot{T}(0) }_{\text{random force}, f^R(t)}.
\end{equation}
Here, $\mc{Q}=\mathbb{1}-\mc{P}$, and we have suppressed possible dependencies on the observables, $\mc{K}(t-s; T(s), \dot{T}(s)) \approx \mc{K}(t-s)$. The first term on the right-hand side corresponds to the effective force that a `particle' experiences along the chosen coordinate $T(t)$.
\footnote{
In the original problem appropriate to Langevin dynamics, the probability is uniform because the Brownian particle is free; generally, particles move in a potential field that influences the dynamics in a Markovian way (see Appendix~\ref{app:GLE})
} 
To calculate the effective force we need the histogram of the time series, specifically the quantity $-\log\mathbb{P}(T)$, and the mass, $M(t) \propto 1/\langle \dot{T}^2\rangle_{T(t)}$, where the subscript denotes this as a conditional average where we only count observations of $\dot{T}^2$ in the time series where $T$ takes a particular value $t$ steps ago. When statistics are Gaussian, $\mathbb{P}(T)\sim\exp(-aT^2)$ and so the `potential of mean force' (PMF) is parabolic. Often, the mass is a constant, but not generally: if $\dot{T}^2$ depends on $T(t)$, the effective force differs from the mean force. The second term describes the memory contribution. 
We extract the kernel $\mc{K}(t) = \langle \dot{T}\mc{LQ}\e{\imi\mc{QL}t}\mc{QL}\dot{T}\rangle / \langle \dot{T}^2\rangle$ using a numerical recipe containing correlation functions of two-point correlations $\langle X(t)Y(t+s)\rangle$ for $X,Y \in \{T, \dot{T}, \ddot{T} \}$ spanning the temperature and its first two time derivatives. 
The angular brackets denote the ensemble average, which we evaluate with a sliding window over the trajectory for an equilibrium series. The final noise term is related to the memory kernel via the equilibrium fluctuation-dissipation theorem, but for our purposes we do not need to construct the noise term explicitly. For further exposition, including the different projection operators $\mc{P}$ compared below, see Appendix~\ref{app:GLE}.

We begin by following the protocol in Ref.~\onlinecite{Kiefer2025} to preprocess publicly available Boulder temperature time series data from Feb.~1992 to Oct.~2025.\cite{NOAA_Boulder} Figure~\ref{fig:1a} displays a fragment over 4~years, illustrating that fluctuations in winter months (blue) are larger than those in summer months (red). Figure~\ref{fig:1b} shows that the histogram of the series is slightly bistable with a pronounced cold tail. The differences in magnitude of the fluctuations suggest the data are not time-translationally invariant. We then filter the annual and zero frequency parts, recovering fluctuations about the deterministic mean,
\footnote{The data reveal a warming equivalent to around 0.5 degrees per decade.} 
$\mathbb{B}_{DC}[\mathbb{B}_{1/365}[T(\omega)]]=T_f(\omega)$, as shown in Fig.~\ref{fig:1c}. Unlike the data from Berlin and Durham, the histogram of Boulder's filtered data is not Gaussian. Nevertheless, summer months give a sub-population that is fairly symmetric and Gaussian, with slightly fat tails. In contrast, winter months exhibit an asymmetric distribution with a thin tail to the hot side and a thick tail to the cold. 

\begin{figure}[!t]
    \centering
    \includegraphics[width=0.9\linewidth]{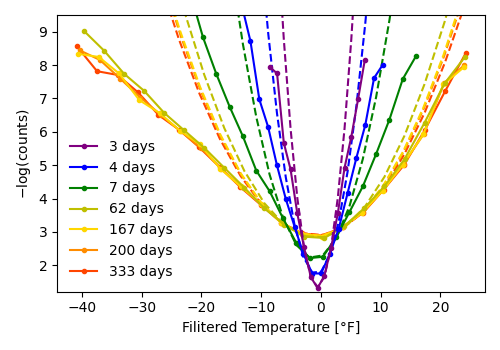}
    \caption{Histograms resulting from progressively larger highpass filters. Dashed lines are quadratic fits to the points around the minima. Non-Gaussianity is reduced but ultimately persists until the shortest timescales.}
    \label{fig:heavy_filtering}
    \vspace{10pt}
\end{figure}

\begin{figure*}[!t]
    \centering
    \begin{subfigure}[b]{0.45\textwidth}
        \resizebox{\textwidth}{!}{\includegraphics{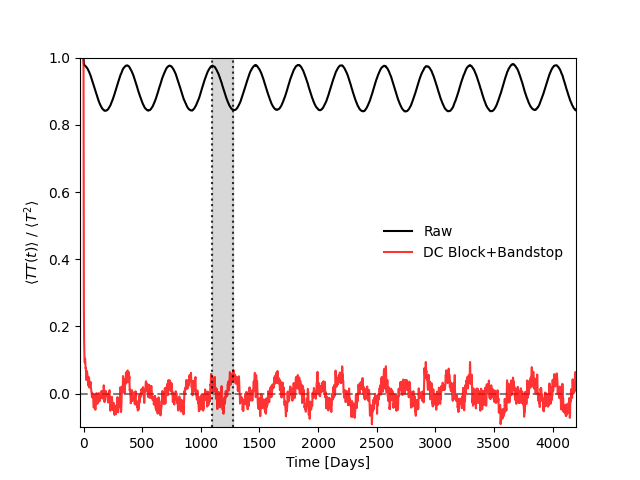}} 
        \caption{}
        \label{fig:2a}
    \end{subfigure}
    \begin{subfigure}[b]{0.425\textwidth}
        \resizebox{\textwidth}{!}{\includegraphics{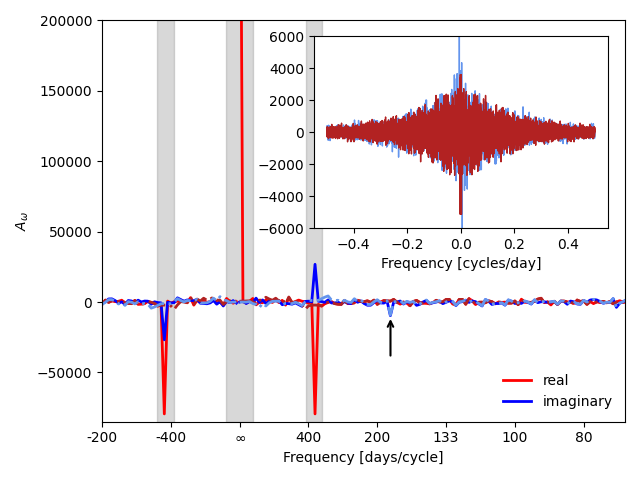}} 
        \caption{}
        \label{fig:2b}
    \end{subfigure}
    \\
    \vspace{-5pt}
    \begin{subfigure}[b]{0.45\textwidth}
        \resizebox{\textwidth}{!}{\includegraphics{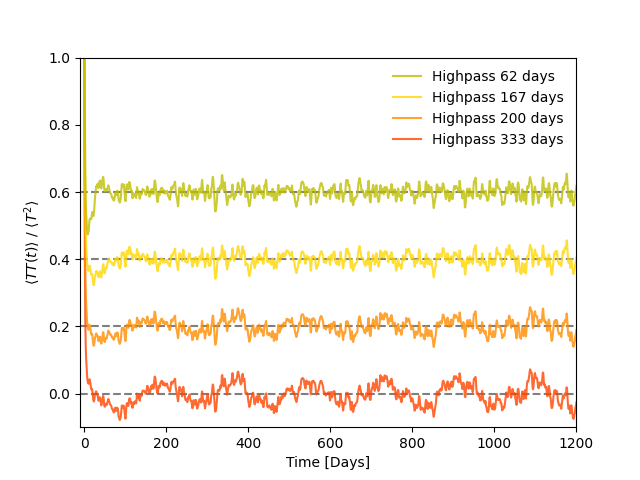}} 
        \caption{}
        \label{fig:2c}
    \end{subfigure}
    \begin{subfigure}[b]{0.425\textwidth}
        \resizebox{\textwidth}{!}{\includegraphics{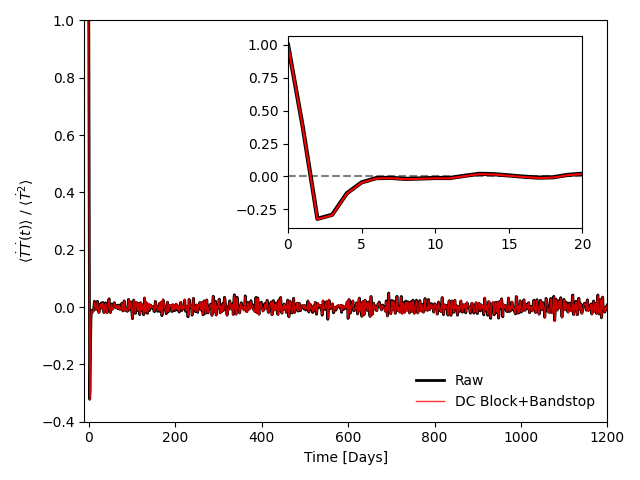}} 
        \caption{}
        \label{fig:2d}
    \end{subfigure}
    
    \caption{Constructing the temperature correlation function. \textbf{(a)} Seasonal data is annually periodic with an initial decay. Filtering in panel (b) (dot-dashed lines) removes the average correlation part but leaves oscillation about zero on a half-year timescale. \textbf{(b)} Filtering of FT data (solid lines) in frequency space at the DC and annual frequencies (gray shading). In this first attempt, we do not remove the first harmonic (which is in the imaginary part); as in Ref.~\onlinecite{Kiefer2025}, this component was not significant. \textbf{(c)} Using a high pass with an increasing cutoff results in changes until reaching 167 days (half a year). Now the harmonic is removed. Further filtering changes the  correlation cage region depth and timescale. \textbf{(d)} Applying a time derivative on the raw and filtered (panel b) data gives the same result since the analytically derived multiplication by $\omega$ suppresses low-frequency contributions. Inset zooms into the early-time dynamics, revealing a lifetime of just under a week.}
    \label{fig:correlations}
\end{figure*}

The necessity of filtering is evident if one applies the `equilibrium' workflow. This is instructive because the artifacts that arise as a result are clear: the correlation function extracted from the unfiltered data oscillates about a non-zero value representing the periodicity of the data (Fig.~\ref{fig:2a}). Hence, the associated memory kernel will never decay to zero. In contrast, the correlation function of the filtered time series decays rapidly, but then exhibits permanent oscillations about zero (Fig.~\ref{fig:2a}). We track this to the presence of a frequency contribution from the harmonic of the annual cycle---a half-year component---to the deterministic evolution. It is purely imaginary (Fig.~\ref{fig:2b}) and maximal in spring and autumn. Employing additional filtering removes these oscillations and shortens the memory kernel lifetime (Fig.~\ref{fig:2c}). However, excessive filtering begins to distort the histogram (see Fig.~\ref{fig:heavy_filtering}). Interestingly, non-Gaussianity is robust to this procedure: it persists down to the most highly filtered data with only fast timescales remaining.

Turning to the lifetimes, we investigate how the non-equilibrium nature of the system is reflected in the correlation functions. We compare the autocorrelation of $v \equiv \dot{T}$ with that of $x \equiv T$. Even without any filtering, the velocity time series yields a short-lived autocorrelation function without oscillations ($\tau_{VV} \sim$~1 week). This is in contrast to $\langle T(0)T(t)\rangle$, where $\tau_{TT} \sim$~2~months at the highest level of filtering, and longer otherwise. Interestingly, $\langle \dot{T}(0)\dot{T}(t)\rangle$ is largely insensitive to the amount of filtering, and we get quantitatively the same result even if we use the raw, unfiltered data (Fig.~\ref{fig:2d}).
\footnote{This is because double differentiation in the time domain is equivalent to applying a quadratic function in the frequency domain $\partial^2/\partial t^2 \rightarrow -\omega^2$, and this has the effect of down-weighting the low-frequency spectrum (removing the DC $\omega=0$ contribution completely). The same observation is made in Ref.~\onlinecite{Kiefer2025}} 
This order-of-magnitude timescale separation is striking, as it suggests that if the slow seasonal change can be decoupled from the fluctuations appropriate to that time of year, a pseudo-equilibrium approach could be applied.

In sum, Boulder's temperature data should not be modelled with a Mori GLE in the manner of Ref.~\onlinecite{Kiefer2025}. While the filtering procedure aims to recover a univariate time series by collapsing the data, Boulder's data reveals seasonal variations, even after filtering. Indeed, even if one had two sub-populations with non-Gaussian distributions that, when added, conspired to give an approximately Gaussian histogram, the resulting model would only work \textit{on average}. If we can introduce flexibility to change between sub-populations, we will better describe the temperature dynamics.\cite{Ge2024, Knoch2019} That is, the GLE needs to be position (temperature) dependent. A question then arises about the form of the driving. For example, Ref.~\onlinecite{Netz2025} assumes that the external field couples to the observable, and the observable only. Contrastingly, for the weather problem, we imagine the temperature probably does not couple to the external forcing alone, and directly not at all. How then should one tackle both problems simultaneously? To make progress toward an answer, we delineate a possible approach if a homoskedastic but position-dependent and non-Gaussian process could be extracted. 

\section{Double Well With Spatially Dependent Friction}\label{sec:double_well}
A simple test case for the GLE consists of describing the Langevin dynamics of a particle on a one-dimensional symmetric quartic potential---a numerically solvable and physically transparent model that we can easily alter. Here, we do not drive the system, but instead focus on position dependence in the random force and friction kernel. We achieve this by having the random force smoothly switch between a lower and higher amplitude,\cite{Bridge2024}
\begin{equation}
    \tilde{f}^R_\mathrm{white}(x) = \left(1 + \frac{3} {1 + \exp(-4 x)}\right) f^R_\mathrm{white},
\end{equation}
where ${f}_{\rm white}^{R}(x)$ is white Gaussian noise. Our motivation is to create different dissipative Kramers regimes in the two wells,\cite{Kramers1940, Melnikov1986, Lavacchi2025} causing the particle to move between the oscillatory (energy-limited) and overdamped (velocity-limited) regimes when it crosses the barrier. After obtaining position and velocity time series (see App.~\ref{app:double_well}), we construct correlation functions like those we built using the temperature data series in the previous section. 

\begin{figure*}[!ht]
    \centering
    \begin{subfigure}[b]{0.4\textwidth}
        \resizebox{\textwidth}{!}{\includegraphics{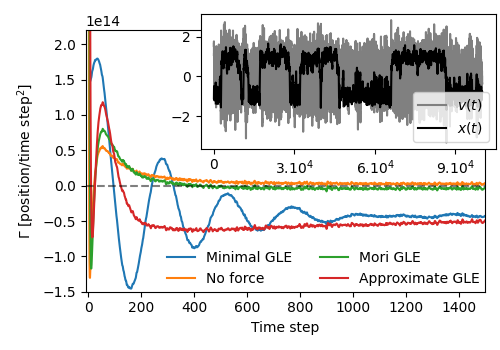}} 
        \vspace{-15pt}
        \caption{}
        \label{fig:3a}
    \end{subfigure}
    \begin{subfigure}[b]{0.4\textwidth}
        \resizebox{\textwidth}{!}{\includegraphics{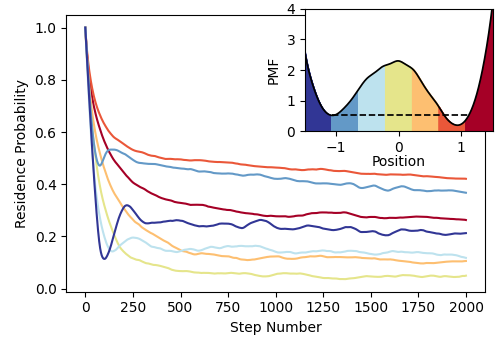}} 
        \vspace{-15pt}
        \caption{}
        \label{fig:3b}
    \end{subfigure}
    \\
    \vspace{5pt}
    \begin{subfigure}[b]{0.4\textwidth}
        \resizebox{\textwidth}{!}{\includegraphics{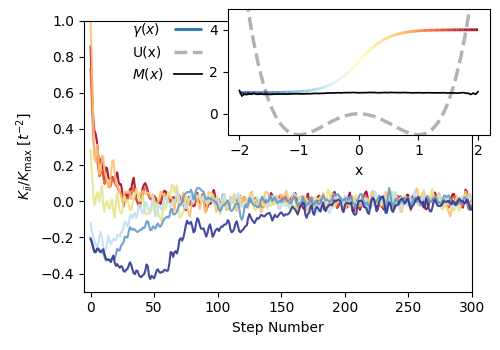}} 
        \vspace{-15pt}
        \caption{}
        \label{fig:3c}
    \end{subfigure}
    \begin{subfigure}[b]{0.4\textwidth}
        \resizebox{\textwidth}{!}{\includegraphics{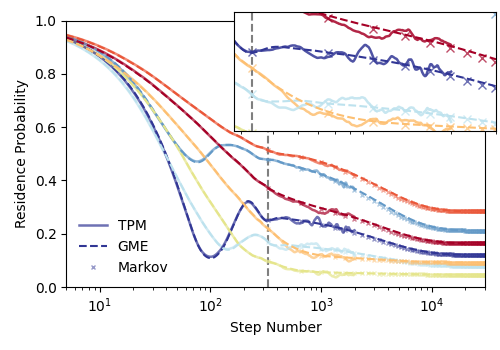}} 
        \vspace{-15pt}
        \caption{}
        \label{fig:3d}
    \end{subfigure}
    \caption{Position-dependent friction for degenerate double-well; the mass is independent of position because the GLE describes a phase-space coordinate. \textbf{(a)} GLE applied directly to the position time series using a variety of definitions for the kernel. Inset: fragment of the trajectory showing the jumps between wells, with larger fluctuations in the left well. \textbf{(b)} Partitioning of the position coordinate into 7 states. Inset: PMF, where the dashed line is a guide to the eye showing asymmetry caused by the friction. \textbf{(c)} Diagonal elements of the memory kernel from the states defined in panel (b). Qualitatively different behavior in the two wells' signals with position-dependent friction in a state-based picture (inset). \textbf{(d)} State-based modelling allows an early memory cutoff GME. Inset: crosses show the Markov limit (i.e., a Markov state model) with lagtime equal to the GME cutoff and resolution set by its lagtime.}
    \label{fig:double_well}
\end{figure*}

One may close the GLE by integrating against $\dot{x}_0$ and averaging over the equilibrium distribution to obtain the velocity autocorrelation function $C(t) \equiv \langle \dot{x}_0\dot{x}_t \rangle$,
\begin{equation}
    \dot{C}(t) = G_i(t) - \int_0^tds\ C(s)\mc{K}_i(t-s; x_s, p_s). 
\end{equation}
Here, $G_i(t)$ depends on the projector (see App.~\ref{app:GLE}). This closed GLE is referred to as a generalized master equation (GME). One can extract the memory function via numerical inversion of the discretized correlation functions,\cite{Sayer2024, Kiefer2025}
\begin{align}
    \mc{LHS}(n\mathrm{\delta t}) &= \sum_{i=0}^{n-1} w_i \mc{K}(n\mathrm{\delta t}-i\mathrm{\delta t}) \mc{RHS}(i\mathrm{\delta t}) \\
    \mc{K}(n\mathrm{\delta t}) &= \frac{2}{\mc{RHS}(0)}\Big[ \mc{LHS}(n\mathrm{\delta t}) \nonumber\\
    &\hspace{30pt}- \sum_{i=1}^{n-1}w_i\mc{K}(n\mathrm{\delta t}-i\mathrm{\delta t})\mc{RHS}(i\mathrm{\delta t}) \Big],
\end{align}
where $w_i = 1-\delta_{0,i}-\delta_{i,n-1}$ are the trapezium rule weights and we engineer the units such that ${\mc{RHS}(0)=1}$. By choosing different expressions for the left- and right-hand sides ($\mc{LHS}$ and $\mc{RHS}$), we recover different momentum-independent kernels (appropriate to Eqs.~\ref{eq:GLE_mori},~\ref{eq:GLE_approx}, and~\ref{eq:GLE_vroylandt}), including one where the potential term, which combines with $\ddot{x}_t$ to form the $\mc{LHS}(t)$, is omitted. For example, the Mori projector (Eq.~\ref{eq:GLE_mori}) has
\begin{align}
    \mc{LHS}(t) &= \left\langle \dot{x}_0\left(\ddot{x}_t+\Omega(x_t-x_{eq})\right)  \right\rangle, \\
    \mc{RHS}(t)&=\langle \dot{x}_0\dot{x}_t\rangle,
\end{align}
while the minimal GLE (Eq.~\ref{eq:GLE_vroylandt}) has
\begin{align}
    \mc{LHS}(t) &= \left\langle \dot{x}_0\left(\ddot{x}_t+\nabla_x\Phi_\mathrm{MF}(x_t)/M\right)  \right\rangle, \\
    \mc{RHS}(t)&=\langle \dot{x}_0\dot{x}_t\nabla^2\Phi_\mathrm{MF}(x_t)\rangle.
\end{align}
We show the friction kernels in Fig.~\ref{fig:3a} and a fragment of the trajectory in the inset. GLEs here have memory kernels with long lifetimes, as is often found.\cite{Lange2006, Carof2014a, Lesnicki2016, Kowalik2019, Bruenig2022, Vroylandt2022a, Vroylandt2022b, Kiefer2025b, Kiefer2025, Vroylandt2022c} We trace this to $\langle \nabla_x\Phi_\mathrm{MF}(x_0)\dot{x}_t\rangle$---which for our potential is close to $\langle x_0 \dot{x}_t\rangle$---having a particularly long decay time. Yet, we know by construction that kernels properly varying in position would be delta-like, as the noise source is white. It is quite unhelpful that a local variability in friction should give rise to memory that persists on the timescale of transitions between wells, and suggests an alternative approach ought to be more effective.

In biophysics, diffusive barrier-crossing processes are ubiquitous.\cite{Jacob1999, Best2005, Best2006, Best2010, Neupane2016, Ayaz2021} One of the most celebrated approaches to tackle the problem is to construct a projector that classifies the states of the system:\cite{Zwanzig1983, Hummer2014} a protein might be unfolded, folded, or in some other misfolded or partially folded configurations. While finding the optimal definition for these states remains an open question, the idea is to remove the multitude of ``unimportant'' fast motions of the structure and concentrate on slow conformational switching. Simulation data can then be used to build the transition probability matrix (TPM), which is the conditional average
\begin{equation}
    {\mc{C}}_{ij}(t) = \mathbb{P}\left( \chi(t) \in V_j ~\vert~ \chi(0) \in V_i \right)
\end{equation} 
for time-dependent indicator functions $\chi$ and volume $V_s$ including all points that map to state $s$. One commonly discards velocity and only builds states (or collective variables) from positions. This dynamical matrix is then the target output of the model. The discretized integrated GME, the transfer tensor method (TTM),\cite{Cerrillo2014a} is
\begin{equation}\label{eq:qMSM}
    \bm{\mc{C}}(n\mathrm{\delta t}) = \sum_{i=0}^{N-1}\bm{\mc{K}}(n\mathrm{\delta t}-i\mathrm{\delta t})\bm{\mc{C}}(i\mathrm{\delta t}),
\end{equation}
where the elements of the memory kernel $\bm{\mc{K}}(t)$ quantify the history dependence of the transition probability. If ${\bm{\mc{K}}(i\mathrm{\delta t})=0}~\forall~i>1$, the dynamics are Markovian and the resulting equation is that of a Markov State Model (MSM).\cite{Chodera2014, Husic2018} The key point is that Equation~\ref{eq:qMSM} tacitly includes position-dependence because the rates of moving between different states are separate quantities, each associated with a different friction component. Allowing different equations of motion for each pair of states is therefore the state-based analog to having a position-dependent GLE kernel.

To apply this framework to our double-well system, we partition the 1D position coordinate into states. We choose a small number of equally spaced intervals with open endpoints. Regularity is not necessary, and the refinement procedure for choosing bins to optimize model performance has been carefully studied.\cite{Husic2018} However, the available statistics constrains the spatial resolution (total number of states) because one needs to converge transition probabilities from the available simulation or measurement data---a feat that becomes increasingly difficult as the number of states grows. Figure~\ref{fig:3b} displays the resulting residence probabilities ($C_{ii}(t)$), which quantitatively capture the differing dynamics of the two wells that we qualitatively understand from observing the trajectory. With this time-dependent matrix, we can invert Eq.~\ref{eq:qMSM} to obtain $\bm{\mc{K}}(t)$. Figure~\ref{fig:3c} shows its diagonal elements, which change qualitatively as one moves between left and right wells. Indeed, these elements have both magnitudes and lifetimes that depend upon their position. The longest lifetime is that of the left-most element with the lowest friction ($\sim$~200~timesteps), but this is still shorter than any GLE in Fig.~\ref{fig:3a}. Since the memory kernel determines the TPM, its long-time values, (equilibrium populations) are accessible by evolving Eq.~\ref{eq:qMSM} with a depth of fewer than 200~data points. We present this extension, with comparison to simulation data and corresponding Markov approximation with $\mathrm{\delta t}\rightarrow \tau_\mc{K}\mathrm{\delta t}$, in Fig.~\ref{fig:3d}. Even with such few states, the Markov approximation gives good accuracy. The TPM-GME therefore captures the different strengths of the friction in each well in a completely unsupervised fashion.

\begin{figure*}[!th]
    \centering
    \begin{subfigure}[b]{0.2\textwidth}
        \resizebox{\textwidth}{!}{\includegraphics{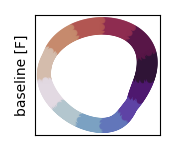}} 
        \\
        \resizebox{\textwidth}{!}{\includegraphics{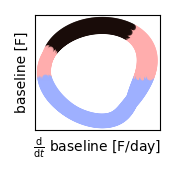}} 
        \caption{}
        \label{fig:4a}
    \end{subfigure}
    \begin{subfigure}[b]{0.1\textwidth}
        \resizebox{\textwidth}{!}{\includegraphics{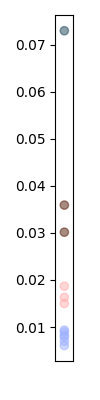}} 
        \vspace{-12pt}
        \caption{}
        \label{fig:4b}
    \end{subfigure}
    \begin{subfigure}[b]{0.575\textwidth}
        \resizebox{\textwidth}{!}{\includegraphics{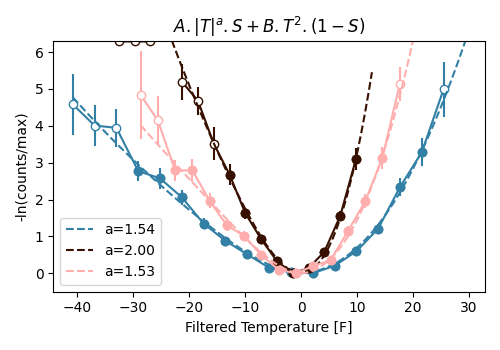}} 
        \caption{}
        \label{fig:4c}
    \end{subfigure}
    \caption{\textbf{(a)} The baseline temperature series, $T_b$ (i.e., the part removed by the filter) and its time derivative $\dot{T}_b$ mapped to normalized polar coordinates. We split this map into regularly spaced (angular) states (Top), which we then aggregate into seasons based on clustering (Bottom). \textbf{(b)} K-means clustering is performed on the $A$ parameter resulting from a fit of Eq.~\ref{eq:asymmetric_histogram} to the filtered temperature histogram conditioned on each regular state; visually, we choose $k=4$; we then include the outlier with its nearest cluster to have sufficient statistics, resulting in 3 clusters in total. \textbf{(c)} Histograms conditioned on each aggregated state or `season'. Open circles have $n<30$ observations and are not included in the fit; error bars are $2\sigma/\sqrt{n}$.}
    \label{fig:3d_states}
\end{figure*}

How can we understand the discrete TPM-GME being more efficient than the (in principle, exact) GLE? By choosing a basis of indicator functions rather than delta functions, we sacrifice some resolution. Indicator functions with large volumes lead to longer memory times, as one combines motions with short and long timescales and thus offloads slow modes into the complementary space that defines the memory kernel. In this limit, a MSM works poorly because equilibration within a state is slow compared to transitions between them.\cite{Prinz2011} In contrast, using indicator functions with small volumes approaches the Markovian limit (the principle behind microstate aggregation in MSM-building workflows). But does it also simultaneously approach the GLE limit? As an illustration, we binned our double well into between 2~and~49 states, and found that the memory lifetime monotonically decreased, suggesting that the GLE should have an even shorter lifetime for delta-like indicator functions. However, this cannot be verified in the TPM-GME framework with finite sampling because the data eventually become insufficient with an increasing number of states, as many bins do not see transitions between one another. 

This reveals an efficiency tradeoff between the GLE and TPM-GME in the limit of a stationary time series with Gaussian fluctuations. When the memory kernel is position-independent,\cite{Ayaz2022, Vroylandt2022b} the GLE becomes more efficient because converging the memory kernel benefits from the self-averaging of the entire trajectory and does not require seeing transitions between all parts of the space. In such cases, all that remains is to converge the PMF. As a one-time quantity, the PMF is generally easier to converge than the time-interval dependent TPM, where one must converge with respect to interstate transitions. When the memory kernel is position-dependent, the TPM-GME is more efficient because it encodes the position-dependence of the memory kernel in its state aggregation at no additional cost. That is, the GLE memory kernel becomes an infinite-dimensional matrix in the basis of positions, which requires converging the transitions between states in this basis. In this latter case, one can imagine the state-based TPM-GME approaching a \textit{position-only} GLE in the limit of delta-function states. 

\section{Floquet drive and homoskedasticity}\label{sec:folding}
We have deduced that direct application of the TPM-GME approach to our weather data would not work: if we apply it to the filtered data, $T_f(t)$, we would lose all our position (temperature) dependence, as the original GLE approach would; if we apply it to the raw data, $T(t)$, our description would be dominated by the steady-state drift of the forcing, preventing us from obtaining a reasonable memory equation (see Fig.~\ref{fig:FTPM}). What is desired is a way to separate the position in the yearly cycle from the fluctuations about that mean---a proper, mathematically-rooted definition for different seasons.

How then does one assign a particular fluctuation's yearly position? Filtering returns fluctuations, but also a residual baseline
\begin{equation}
    T_b(t) = T(t)- T_f(t).
\end{equation}
The central insight of Floquet theory is that dynamical variables (probability density, transition matrix, etc.) under a periodic drive of period $S$ also exhibit the same period. Deviations from this periodic steady-state exist transiently at the start of the motion, but after relaxation, all information repeats within a single period,
\begin{equation}
    T_b(t+S) = T_b(t).
\end{equation}
For climate data, it is tempting to assign $S \approx 365$~days. However, $S$ is the lowest common multiple of all periodic components, of which there are many, including much longer climate cycles from the Biennial oscillation of stratospheric winds and El Ni\~no/La Niña all the way up to geological and even stellar timescales.\cite{Ghil2020, Todd2025} Thus, folding the total trajectory onto a yearly cycle produces week-to-week histograms that have visible, shifted sets of outliers, consistent with data being locally incommensurate with a single-frequency motion (Fig.~\ref{fig:weekly_series}). This is the manifestation of the complex beating evident under the fluctuations (Fig.~\ref{fig:1c}). A more agnostic way to fold the trajectory is to determine its correspondence to the filtered, baseline-driven cycle $T_b(t)$, which captures deterministic local frequency variations. By plotting all points as 2D coordinates of $T_b(t)$ and $\dot{T}_b(t)$ (Fig.~\ref{fig:4a}), we get smooth, continuous circuits with no outliers and an annual thickness representing the range of amplitudes that compose the total time series.
\footnote{The width of the circle represents the range of variations arising from multiple driving frequencies. If there were just one frequency, the cycle would be a single point wide.} 
Since the goal is to describe correlated temperature fluctuations linked to the time of year, we leverage this $T_b$ spirograph to separate seasonal temperature fluctuations based on their statistical distributions. If this is possible, we will be able to construct a TPM-GME for each independent set. 

In the extreme case, one might imagine building a TPM quantifying the transition probability between a given day and its specific temperature fluctuations into subsequent days. This amounts to the delta-function TPM-GME, which would have prohibitively large data costs. To maintain a balance between the number of states and data availability, and because we suspect certain months will behave similarly (the `summer' and `winter' from our intuition), we discretize our 2D coordinates into a small number of states that may display different statistics but exhibit time-translational invariance.\cite{Knoch2019} Proceeding as in the MSM workflow,\cite{Husic2018} we start with a relatively large number of 12 `micro' states. To build these, we choose a regular, angular partitioning for simplicity, which is shown as a 12-color map in Fig.~\ref{fig:4a}-top. The maximum number of states is partly determined by the amount of reference data one can use to build the TPM-GME for the particular problem. We then aggregate these into just 4 `macro' states---the mathematically derived `seasons'. For this, we fit the histogram conditioned on each microstate to the asymmetric PMF,
\begin{equation}\label{eq:asymmetric_histogram}
    \mathbb{P}(T_f) = A|T_f|^a\Sigma(T_f) + BT_f^2(1-\Sigma(T_f)),
\end{equation}
where $\Sigma$ is a sigmoid function to ensure smoothness, and then perform k-means clustering on the $B$ parameter. In fact, besides one outlier, we find the data aggregate into three rather than four seasons (Fig.~\ref{fig:4b}). Physically, this merges the 3 hottest months as `summer', the 6 coldest months as `winter', and 1~month in spring with 2~months in autumn for a combined `equinoctial' season, as shown in Fig.~\ref{fig:4a}-bottom. Thus, although there is some correspondence, the mathematical seasons of similar fluctuations differ from both equally divided calendar seasons and orbitally-derived astronomical seasons. 

We reanalyze the combined histograms of these three seasons and present them in Fig.~\ref{fig:4c}. Again, we use the asymmetric form Eq.~\ref{eq:asymmetric_histogram}, which accurately fits the data in all cases. While the summer grouping has an asymmetric histogram, it is nevertheless quadratic on both sides. The other two seasons have pronounced stretched-exponential character to the cold: in both cases $a\simeq 1.5$ but with different prefactors. This aggregation of states with equivalent histograms constitutes our \textit{homoskedastic projection}. We note, however, that while time-translational invariance implies homoskedasticity, the converse is not guaranteed. We nevertheless introduce this homoskedastic projection in the hopes of identifying pseudo-stationary regions that permit one to easily build a TPM-GME.

\begin{figure*}[!th]
    \centering
    \begin{subfigure}[b]{0.3\textwidth}
        \resizebox{\textwidth}{!}{\includegraphics{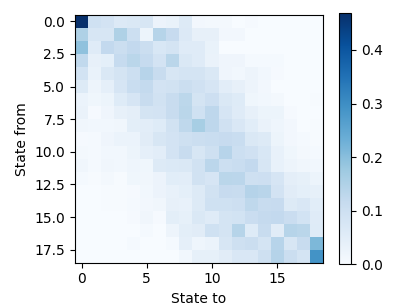}} 
        \caption{}
        \label{fig:5a}
    \end{subfigure}
    \begin{subfigure}[b]{0.3\textwidth}
        \resizebox{\textwidth}{!}{\includegraphics{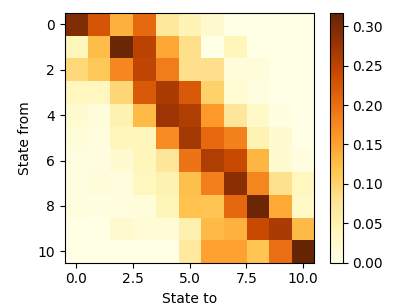}} 
        \caption{}
        \label{fig:5b}
    \end{subfigure}
    \begin{subfigure}[b]{0.3\textwidth}
        \resizebox{\textwidth}{!}{\includegraphics{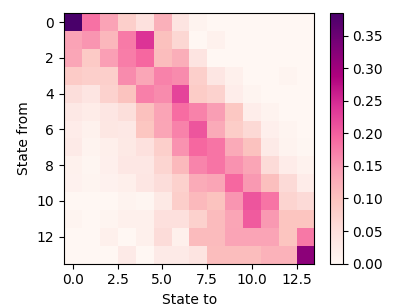}} 
        \caption{}
        \label{fig:5c}
    \end{subfigure}
    \\
    \hspace{-18pt}
    \begin{subfigure}[b]{0.45\textwidth}
        \resizebox{\textwidth}{!}{\includegraphics{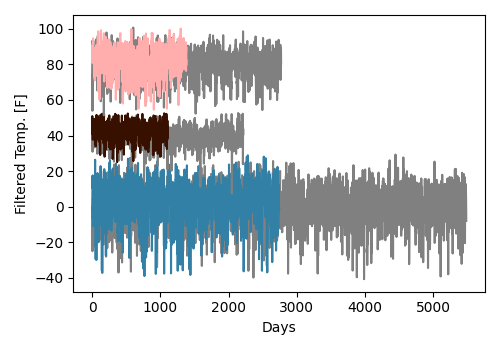}} 
        \vspace{-12pt}
        \caption{}
        \label{fig:5d}
    \end{subfigure}
    \begin{subfigure}[b]{0.45\textwidth}
        \resizebox{\textwidth}{!}{\includegraphics{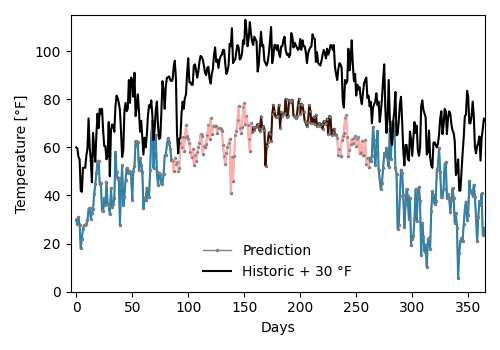}} 
        \vspace{-12pt}
        \caption{}
        \label{fig:5e}
    \end{subfigure}
    \caption{\textbf{(a)--(c)} Markov stochastic matrices at fixed 2~°F resolution for winter (blue), summer (brown), and equinoctial (pink) seasons. \textbf{(d)} $T_f(t)$ time series generated via Markov chain unraveling for each season, half-overlaid with the true data to show successful generation of colored, non-Gaussian noise. \textbf{(e)} Example of a fully predicted time series of baseline plus fluctuations, including changing between different seasons (colored lines). We displace historic data by 30~°F (in black) to facilitate comparison.}
    \label{fig:3d_TTM}
\end{figure*}

\section{Seasonal transition models and weather prediction}\label{sec:final_model}
Our proposed approach to building a dynamical model of weather (temperature) fluctuations leverages exact and reversible steps to remove the effect of the driving and identify homoskedastic `seasons'. Now, our projection becomes the basis of the first approximation in our treatment: we assume that each (connected) segment of the total time series within a single season is a fragment of a pseudo-stationary trajectory. This enables us to construct pseudo-equilibrium correlation functions for each segment, and then average over segments. To blend the models across seasonal changes, we would need to compare a zero-correlation renewal process with a mixed, inter-state memory. However, anticipating our results, we find all our GMEs satisfy Markovianity, obviating the need for a separate, inter-season parameterization. While, in principle,\cite{Kwasniok2013, Ge2024} each season could have its own GLE, we show that a TPM-GME is simpler and more efficient for weather predictions. We illustrate our protocol, its performance, and benefits below.

\begin{sidewaysfigure}[!ht]
    \includegraphics[width=\linewidth]{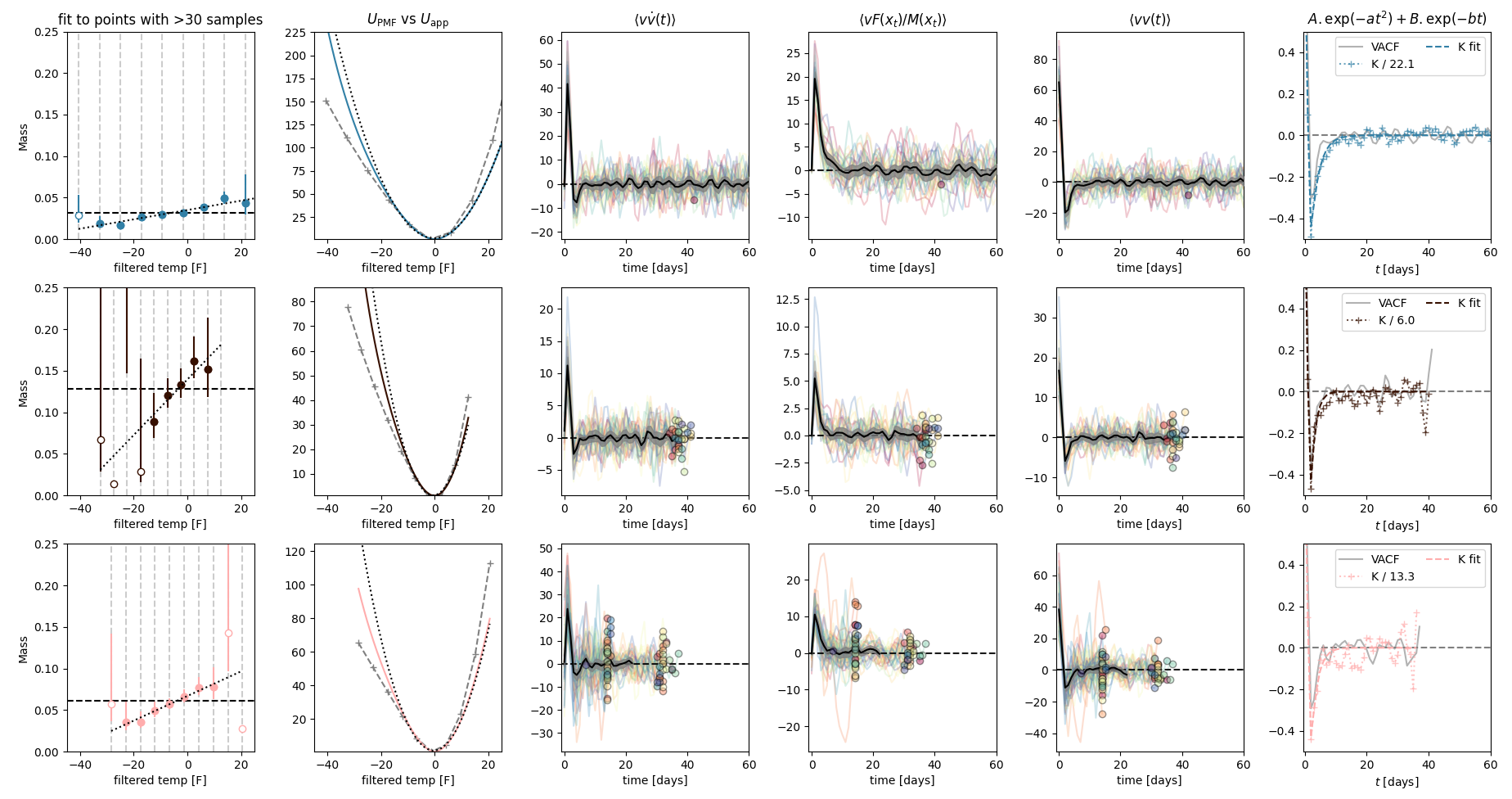}
    \caption{The approximate GLE with variable mass. Each row represents one of the three seasons found by clustering. \textbf{Column 1}: position dependent mass with linear fits, 2 sigma error, empty points not included. \textbf{Column 2}: Apparent potential from Eq.~\ref{eq:GLE_zwanzig_mass} compared to just the PMF; dotted line is quadratic fit to minimum for reference. \textbf{Columns 3-5}: Correlation functions needed for the GLE; coloured lines are individual trajectories, with circle to show their end-time (each year the seasons aren't the same length, and the equinoctal season is split each year into two uneven lengths). \textbf{Column 6}: Memory kernel with two-component fit, overlaid with VACF (column 5); values are normalized for visual comparison and the factor is given in the legend.}
    \label{fig:GLE}
\end{sidewaysfigure}

Figure~\ref{fig:GLE} shows the temperature correlation functions needed to construct a GLE with position-dependent mass for summer, winter, and equinoctial seasons. Fascinatingly, although the seasons have different masses and PMFs, the position-dependence of the mass renders the `apparent' potentials more symmetric---almost Gaussian (Fig.~\ref{fig:4c}). This sheds some light on how the Mori GLE, where position-dependences are absent and the PMF replaced by a Gaussian, can perform surprisingly well (see Appendix~\ref{app:GLE-app-pot}). Although the magnitudes of the kernel vary between seasons, in each case, the functional form is reasonably well fit by a delta-like term and an exponential tail, as in Ref.~\onlinecite{Kiefer2025} (although the sign of our exponential is flipped). As with many GLEs, the memory kernels appear to live longer than the correlation functions they encode, although not by much: for every season, both functions decay on the timescale of around one week, similar to the global velocity autocorrelation function (see Fig.~\ref{fig:2d}).

Motivated by our observation in Sec.~\ref{sec:double_well} that a state-based analysis can reduce the memory lifetime compared to the GLE, we now pursue a TPM-GME description. In contrast to Sec.~\ref{sec:double_well}, the temperature GLE exhibits position dependence of the mass. While we can capture this dependence in our GLE of Fig.~\ref{fig:GLE} by proposing a functional form that holds for the observed statistics, this requires intuition and user input. Moreover, we neglect the position dependence of the GLE's memory kernel \textit{within} each season, capturing only differences between seasons. We overcome these limitations by adopting the TPM-GME, which automatically includes position dependence in the memory kernel. The cost is loss of resolution. For simplicity, we set a 2~°F temperature resolution across our intra-season models, but note that this is a particular choice. In fact, as the seasons have both different ranges for their fluctuations and different durations, they are not equally sampled, and so do not afford the same level of convergence: our choice of 2~°F represents a more challenging resolution for the high-variance winter season than the other lower-variance seasons. Ultimately, the resolution is determined by the needs of the researcher and the practicalities of the sampling.  

We can now construct the TPM-GMEs for the temperature dynamics within the three seasons. The intra-seasonal TPM is statistically noisy, and most off-diagonal elements are small with only a few percent probability. Hence, we extract a similarly noisy memory kernel. Yet, a Markovian truncation of the memory to one step recovers predictions that quantitatively capture the exponential decay and lifetime of the TPM (Fig.~\ref{fig:winter_TPM}). Later cutoffs are progressively worse, as the kernels are delta-like and statistical noise quickly dominates. We are therefore justified in reducing the GME to a MSM. Figure~\ref{fig:5a}-\ref{fig:5c} presents the Markov stochastic matrices for each season, which have a banded, asymmetrical form. Although the most likely events reside in the same state or transition to a neighboring state (i.e., a change of less than 4~°F), large day-to-day swings are observed and sometimes with reasonable probability. Our TPM-GME's Markovianity mirrors the GLE's strong, delta-like component of the memory, but is unencumbered by an exponential tail. The result is convenient because the propagation of trajectories (rather than average probabilities) required for numerical weather prediction in the presence of non-Markovian effects is involved.\cite{Zwanzig1983, Diosi1997, Zhang2019}

Since the stochastic matrix fully describes a MSM, we can now predict all possible fluctuation time series, and compare them to the historic result. From one step to the next, the matrix dictates the probability of staying in the current temperature bin, or of making a transition. One can unravel this probabilistic evolution into a trajectory by stochastically choosing \textit{one} transition by drawing a random number from a uniform distribution spanning $[0, 1]$ and aligning the result with the weighted conditional probabilities for the current state. This yields a time series with instantaneous values at 2~°F resolution, except for the end bins where the value is not well-defined. This is because the available data does not (and can never) support equal subdivisions of the entire range $T \in (-459.67, \infty)$~°F. That is, the end bins are half-open (see Fig.~\ref{fig:sub_histograms}a). As such, the temperature ranges are much larger within these end bins, encompassing extreme (albeit exponentially improbable) values.

The large temperature ranges of the end bins are a statistical requirement set by the available transition statistics. Yet, while obtaining adequate 2-point quantities to parametrize the TPM-GME necessitates a coarse projection, we still have access to the original data and the associated 1-point histogram. The 1-point quantity is, for the same sampling, much better converged and can therefore support a finer grain. Hence, we can generate a high-resolution twin of our unravelled TPM-GME trajectory by stochastically drawing specific values with sub-2~°F resolution from our previously interpolated distribution (see Eq.~\eqref{eq:asymmetric_histogram}), whose shape we already know can be captured by the stretched exponential with $1<a\leq2$. This hierarchical protocol draws values from each bin's fitted histogram and is completely decoupled from the bin-to-bin TPM-GME. As such, our resolution enhancement does not affect the coarse-grained accuracy of the trajectory, but neglects correlations between transitions associated with these high-resolution values. By doing this, our protocol retains the correct probabilistic evolution of the temperature at the above-2~°F scale, while effecting a sort of mean-field approximation to the `high-frequency' temperature correlations. 

\begin{figure}[!b]
    \centering
    \includegraphics[width=0.9\linewidth]{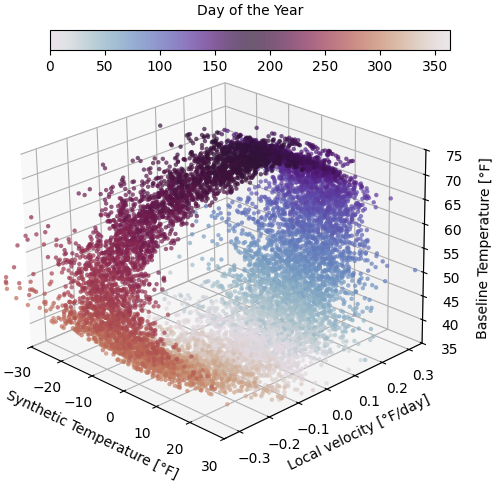}
    \caption{3D scatter plot showing the full time-dependent histogram of the predicted data series in Fig.~\ref{fig:5e}. By construction, the histogram of the predicted data matches the histogram of the historic (reference) data, Fig.~\ref{fig:bagel}.}
    \label{fig:vegan_bagel}
\end{figure}

With this hierarchical protocol in hand, we predict time series for each of the three equilibrium models. Figure~\ref{fig:5d} presents our predicted time series for each season, which visually recapitulate the correct colored, non-Gaussian noise, including the magnitude of tail events. The Markovianity of the TPM-GME is a powerful simplification: all that is needed when the baseline temperature transitions between seasons is for the current state to be reclassified according to the new season's model. We show one multi-season realization in Fig.~\ref{fig:5e}. Of course, predicted trajectories are stochastic, decorrelating in time from a shared initial condition. Yet, the shape of the fluctuations, on the scale of around a week, is evident in both cases: structurally similar features appear in both the historic and our predicted trajectory. At the global histogram level, we can assess how well our modelling steps recover the underlying distributions: Figure~\ref{fig:vegan_bagel} presents the 3D plot of baseline, time-derivative of the baseline (together defining the steady state), and the filtered temperature. Comparison with Appendix Fig.~\ref{fig:bagel} reveals close agreement. Thus, our method successfully preserves the histogram to a high degree of accuracy.

\section{Discussion and Conclusions}
\begin{figure}[!b]
    \centering
    \hspace{-15pt}
    \includegraphics[width=1.05\linewidth]{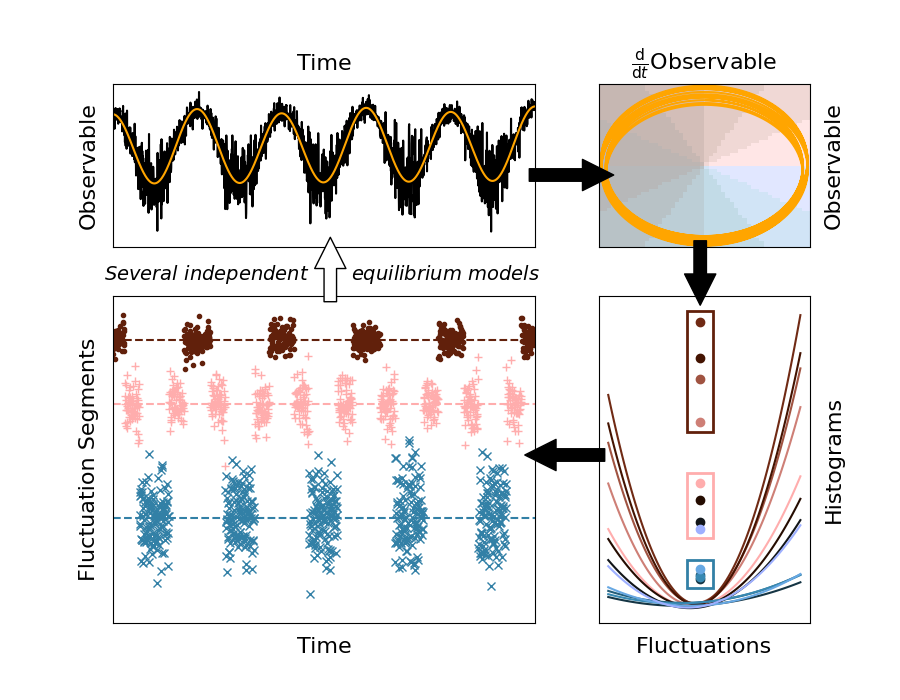}
    \caption{Outline of the overall workflow clockwise from top left. \textbf{Top left:} Begin with a periodically driven time series suspected of having multiple heteroskedastic underlying processes that contribute based on their position in the cycle. \textbf{Top right:} Extract a baseline time series (orange line) using one's choice of filtering (e.g., Fourier), and split the trajectory into microstates. \textbf{Bottom right:} Cluster the microstates based on the histograms of their fluctuations (dots show the curvatures of the histograms, to scale). \textbf{Bottom left:} Leverage the resulting segments of the fluctuation time series to construct an equilibrium model---a GLE or a TPM-based GME. One can then use the state-based model to produce statistics consistent with the original data.}
    \label{fig:schematic}
\end{figure}

We have therefore demonstrated an approach that can accurately and efficiently describe the stochastic dynamics controlling fast-fluctuating variables in nonequilibrium, driven systems---like weather and climate dynamics. The flow of ideas was thus. When a dissipative system is driven by a slow, periodic force (even a bias), it eventually achieves a nonequilibrium periodic state, ensured by the Floquet theorem. Building on recent works, we employed a (reversible, Fourier) filtering protocol to remove the effective periodic drive. Whereas previous works\cite{Ge2024, Kiefer2025} recovered a fast-fluctuating time series that appeared to be generated by a \textit{stationary Gaussian process} suitable for description by a Mori GLE, we tackled the \textit{more general and complex case} where non-stationarity and non-Gaussianity both remain in the residual fluctuations. Here, non-stationarity arose from a change in the feedback mechanism that governs the local weather system, manifesting as a seasonal heteroskedasticity. It gets cold in the winter, and the Boulder weather fluctuates more when it is cold, \textit{not only because it is cold}, but because it is \textit{winter}. Despite the global inapplicability of the equilibrium approach in such cases, we propose that a statistical measure---local skedasticity---can be used to identify time spans over which stationarity may hold and clarify the true extent of non-Gaussianity. Figure~\ref{fig:schematic} schematically presents this workflow.

To treat the residual non-Gaussianity, we employed the mathematically transparent TPM framework. This approach offers three clear advantages over the GLE. First, the TPM construction subsumes complicated questions of position dependence directly into the projector. While finite resolution entails discarding the highest-frequency correlations, we showed that the TPM projection increases efficiency and decreases the memory lifetime of the resulting dynamical model for both our weather data and our model double well. In systems where collecting additional data to improve statistics is expensive or difficult, this parsimony represents a second advantage. In fact, for the 3-seasons we extracted from the Boulder data, each of the resulting models was Markovian. Remarkably, this Markovianity emerged without any coarsening of the time resolution, in contrast to direct modelling of the global average temperature, where researchers found a reduction in resolution from 1 day to 2 months was necessary.\cite{Lak2025} The third advantage is that our TPM-GME approach also disentangles two complex challenges that often get conflated in the GLE approach: non-Gaussianity and the ability to build a rigorous dynamical model. Our homoskedastic projector provided the means to isolate periods of well-defined statistics---indeed, distinctly non-Gaussian ones, with marked asymmetry and polynomial tails into the cold. Identifying the microscopic and even phenomenological origins of non-Gaussian statistics, which are common in climate science,\cite{Bhagavatula1993, Li2005, Sardeshmukh2009, Wen2013, Miotto2021, Tallakstad2013, Packwood2011, Perron2013, Giorgini2024, Gottwald2025, MartinezV2025} continues to motivate a vibrant field of study. Yet, their inclusion remains an open challenge for the data-driven parameterization of the GLE.\cite{Lei2016, Schilling2022, Vroylandt2022a, Lyu2023, Ge2024} This is because constructing the GLE's random force requires knowledge of only the first and second dynamic moments in the Gaussian limit, but generally an infinite number of such moments for the non-Gaussian case.\cite{vanKampenBook} Instead, the inhomogeneous term in the TPM-GME---i.e., the analog of the random force term---disappears due to judicious choice of the projection operator.\cite{Zwanzig1983} Thus, a data-driven construction of the finite-resolution TPM-GME requires only the first and second dynamic moments, regardless of the nature of the statistics. 

Our work enables us to address an important question: when is it easier to construct a GLE than a GME based on observed data? The data-driven GLE is most accessible in the stationary limit; this allows one to build correlation functions from the translating window method and exploit the equilibrium fluctuation-dissipation theorem to construct the memory kernel. The GLE is most efficient when its memory is position-independent, its random force characterized by Gaussian statistics, and its PMF easily accessible, either through enhanced sampling or appropriate filtering. Gaussian statistics are particularly convenient as they require only the second dynamic moment to determine their statistics, and robust protocols exist to generate their noise trajectories, whether with white or colored noise. When any of these conditions ceases to hold, the GME becomes advantageous. Both the GLE and GME require sufficient sampling of the entire dynamical space to construct the PMF or the TPM from the discretized configurational space, respectively. However, unlike the GLE, the GME requires one to choose an aggregation or discretization scheme. Finally, whereas the GLE is ideal for generating \textit{individual} trajectories of the stochastic evolution of the chosen variable, doing the same with the GME is only trivial in the Markovian limit. Overall, while the GLE yields individual stochastic trajectories that enable a user to capture any dynamic moment of the chosen variable, it is most efficient when the process is stationary and Gaussian. In this limit, the GME offers the same advantages while being best suited to interrogate the statistical properties of correlation functions and nonequilibrium averages. 

In conclusion, our present study introduces a data-driven protocol to predict the local surface temperature dynamics in Boulder, Colorado, but has broad-ranging implications. Our work delineates the assumptions and requirements for constructing accurate and efficient stochastic descriptions of the dynamics of fast-fluctuating variables in many-body, nonequilibrium systems subject to a periodic drive. These problems are challenging to traditional GLE-based descriptions and abound in soft matter, biology, and, naturally, global climate and weather prediction. For the latter, our work furthers the Hasselmann program in a complementary way to modern machine learned approaches\cite{Lam2023, Lucarini2023} and sets the stage for refining both numerical weather prediction and climate modelling based on principled use of existing observations of fast-evolving variables (e.g., temperature, pressure, humidity) around the globe. 

\section{Acknowledgements}
We thank Roland Netz, Henrik Kiefer, and Lucas Tepper for sharing their group's GLE code for us to study. We also thank Veronica Vaida and Adrian Tuck for their comments on the manuscript. T.S.~is the recipient of an Early Career Fellowship from the Leverhulme Trust. A.M.C.~acknowledges the support from a David and Lucile Packard Fellowship for Science and Engineering. 

\section{Methods}\label{app:methods}

\begin{figure}[!h]
    \centering
    \vspace{-15pt}    
    \includegraphics[width=0.45\linewidth]{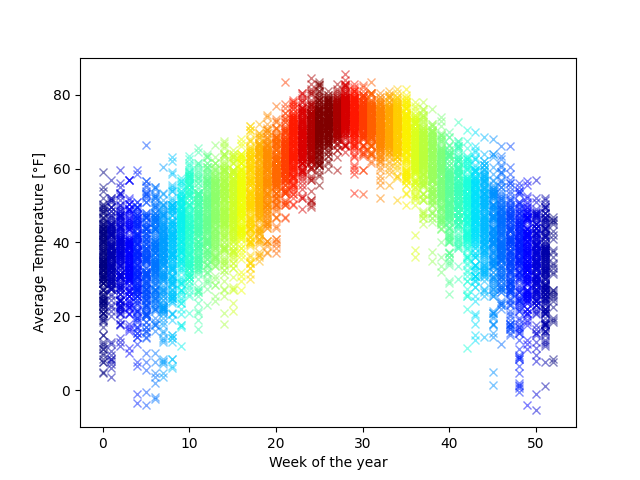}
    \caption{Average temperature data for Boulder, Colorado averaged by week. That is, every week has a data point for every year in the data set. Blue is winter, red is summer. Spread captures fluctuations but also the incommensurate `beating' of the underlying frequencies in the baseline temperature: see Fig.~\ref{fig:3d_states} and text of Section~\ref{sec:folding} for how these are separated.}
    \label{fig:weekly_series}
\end{figure}

\textbf{Filtering:} we used scipy butter filters with values of 
\vspace{-8pt}\begin{spverbatim}
    lowcut = 2.4E-3
    highcut = 1 * 3E-3
    zero_band = 5E-4
\end{spverbatim}\vspace{2pt}
\noindent We followed the workflow of Ref.~\onlinecite{Kiefer2025}. For Boulder, Colorado specifically, we removed the harmonic between  [\verb|0.25*highcut+1.75*lowcut|, \verb|1.75*highcut+0.25*lowcut|], and divided all parameters by the Nyquist frequency of half a day. (see Fig.~\ref{fig:2b}). We discarded the first 2 and last 3 years of the filtered time series, as these display artifacts from the filtering (see Fig.~\ref{fig:1a}). 
Figure~\ref{fig:weekly_series} shows the periodically wrapped, unfiltered data.

\textbf{Histograms:} we used the numpy implementation of doane binning. In Figs.~\ref{fig:dataset}~and~\ref{fig:heavy_filtering}, we applied quadratic fits to the bottom 5 points; in Fig.~\ref{fig:GLE} we extended the fit to any points under 1/4 the maximum value. We fitted the asymmetric function in Eq.~\ref{eq:asymmetric_histogram} for Fig.~\ref{fig:3d_states} with scipy.optimize.curve\_fit.

\textbf{Correlation functions:} we used scipy.signal.correlate and then reweighted the resulting array by the number of samples (i.e., element-wise division by a linear function between 1 and 0.5).

\begin{figure}[!h]
    \centering
    \includegraphics[width=0.45\linewidth]{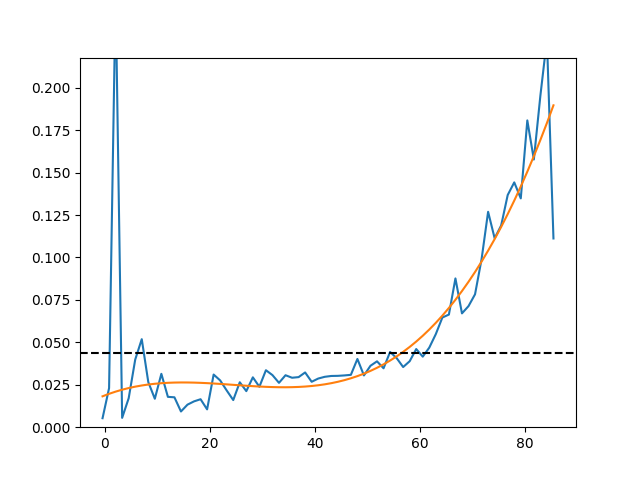}
    \caption{The conditional average of the velocity squared given the position is the mass, and can be position dependent. The solid orange line is a fit to be used in a GLE analysis, and the black dashed line is the average value. When split into states it becomes 3 linear functions, see Fig.~\ref{fig:GLE}.}
    \label{fig:mass}
\end{figure}

\textbf{Conditional averages:} we used the histogrammed series. Hence, for timeseries \verb|fT| histogrammed into \verb|bins|,
\vspace{-18pt}\begin{spverbatim}
lower = np.min(bins)
df = bins[1] - bins[0]
binned_fT = int((fT - lower) // df)
\end{spverbatim}\vspace{4pt}
\noindent We then looped over each bin,computed the averaged quantity $s$ steps in the future, and averaged over the number of samples in that bin. We computed the standard deviation in each bin, and kept track of the number of samples to exclude low-sample (high deviation) values from fits. In most cases, these excluded points still lay on the fit line within their error bars (see, for example, Fig.~\ref{fig:4a}). 
We directly calculated $\langle \dot{T}^2\rangle_{T(t)}$ (see Eq.~\eqref{eq:GLE_zwanzig_mass}). Figure~\ref{fig:mass} shows the mass for the unfiltered data.

\subsection{State Creation and Aggregation}\label{app:methods_states}

When making states from the baseline, we scale to polar coordinates using sklearn
\vspace{-4pt}\begin{spverbatim}
a = vlocal / np.max(vlocal)
b = (baseline) / np.max(baseline)
scaler = StandardScaler()
ab_scaled = scaler.fit_transform(
            np.column_stack((a, b)))
a = ab_scaled[:, 0]
b = ab_scaled[:, 1]
# Convert to polar coordinates
r = np.sqrt(a**2 + b**2)
theta = np.arctan2(b, a)  # angle (-pi, pi)
\end{spverbatim}\vspace{4pt}
\noindent These are the scatter plots in Fig.~\ref{fig:4a}. We then applied K-means with sklearn.

\begin{figure}[!h]
    \centering
    \includegraphics[width=0.45\linewidth]{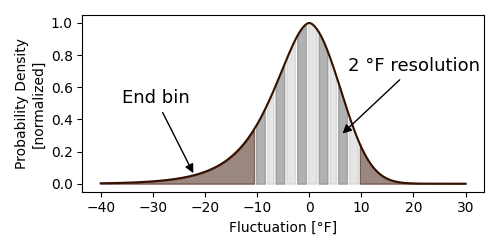} \\
    \includegraphics[width=0.45\linewidth]{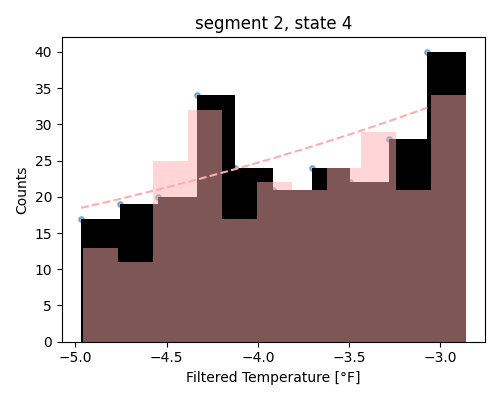}
    \includegraphics[width=0.45\linewidth]{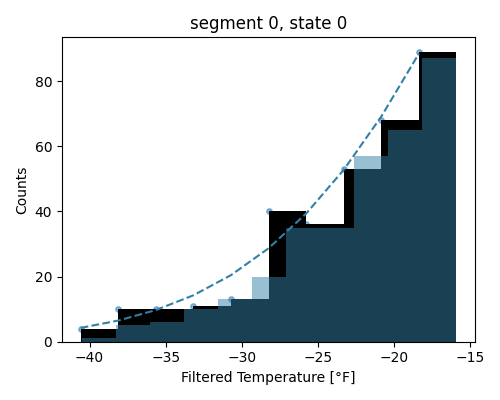}
    \caption{\textbf{Top:} Fluctuation probability density for a season partitioned into 2~°F bins and two end bins. End bins are aggregated from all extreme bins that fall below the threshold, here set as the median number of observations. \textbf{Middle:} Actual and randomly sampled observations for a standard 2~°F bin within the center of the distribution of the equinoctial season, where a roughly-straight line is sufficient to capture the local probability density. \textbf{Bottom:} Actual and randomly sampled observations for an end bin of the winter season, where the probability density is strongly curved, defining an effective width of around 25~°F.}
    \label{fig:sub_histograms}
\end{figure}

To obtain higher resolution, we binned observations in each state again and fitted the stretched exponential, 
\begin{equation}\label{eq:stretched_exp}
    A|T_f|^a + c
\end{equation}
to each (see Fig.~\ref{fig:sub_histograms}). A sufficiently flexible function is crucial because the end-bins require that: 1) a cutoff be achieved so as not to present ultra-extreme values as possible, and 2) more extreme values become progressively less likely, otherwise the histogram of our predicted data would not even qualitatively agree with the historic data on which it was parameterized. Then, when drawing from this distribution, we used
\vspace{-4pt}\begin{spverbatim}
def generic_CDF_choice(func, params, L, H, U):
    xgrid = np.linspace(L, H, 5000)
    pdf_vals = func(xgrid, *params)
    cdf_vals = cumtrapz(pdf_vals, xgrid, initial=0)
    cdf_vals /= cdf_vals[-1]
    inv_cdf = interp1d(cdf_vals, xgrid)
    return inv_cdf(U)
\end{spverbatim}\vspace{4pt}
\noindent where func is the asymmetric Eq.~\ref{eq:stretched_exp}, params come from the fit, L and H are the limits of the bin, and U is the random value drawn at that timestep from the white noise. Figure~\ref{fig:sub_histograms} shows example sample sets.

\subsection{GLE}\label{app:methods_GLE}

Following Ref.~\onlinecite{Kiefer2025}, we extract the memory kernel by numerical inversion of the convolution.  
We do not apply a finite timestep correction. In Ref.~\onlinecite{Kiefer2025} the exponential decay time $\tau$ reduces from 12 days to 3 days as a result of this. For the final weather model with three seasons, we fit each model's memory kernel as a Gaussian plus exponential as
\vspace{-4pt}\begin{spverbatim}
def gamma_yip(t, a, b, tau, tau2):
        gauss_term = a * np.exp(-1 * t**2 / tau)
        exp_term = b * np.exp(-1 * t / tau2)
        return gauss_term + exp_term
        
gamma = gamma_yip
guess = [150, -100, 1, 2]
bounds = ([0, -1500, 0.01, 0.01], [1500, 0, 5, 10])
\end{spverbatim}\vspace{4pt}
Figure~\ref{fig:GLE} shows each term in the GLE. 

\subsection{GME}\label{app:methods_GME}

To locate the memory lifetime, $\tau$, we used an error plot of the L1-norm of predicted versus actual TPM against the choice of upper limit in the convolution (see Eq.~\ref{eq:qMSM}). The error is expected to decrease as a function of increasing $\tau$ until it plateaus. For the double well, this is what we find, and we identify the start of the plateau as $\tau$. 

\begin{figure}[!h]
    \centering
    \includegraphics[width=0.45\linewidth]{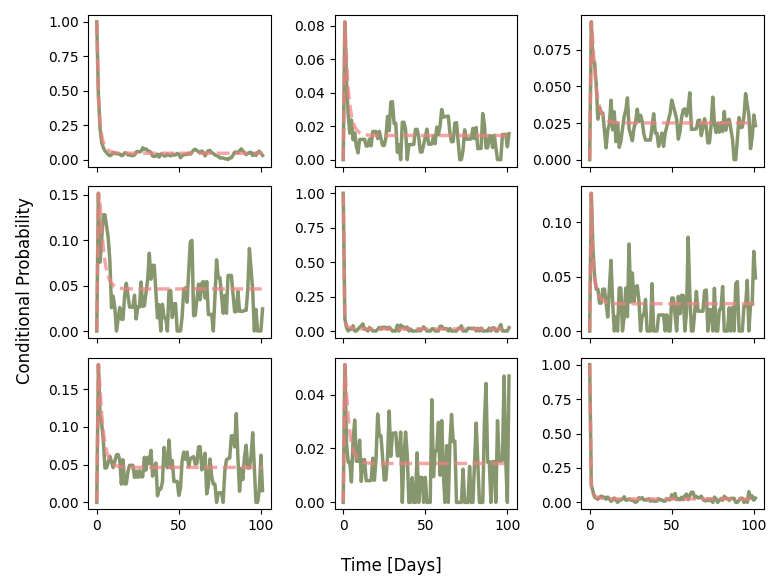}
    \includegraphics[width=0.45\linewidth]{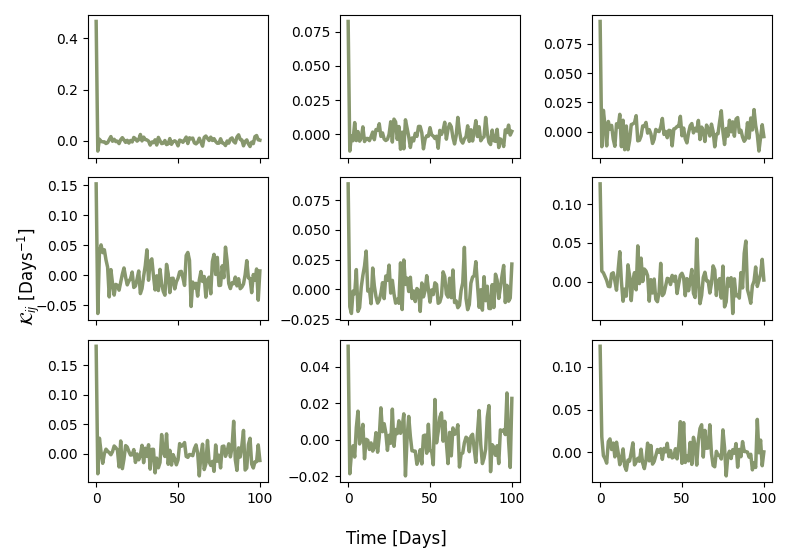}
    \caption{\textbf{Top}: Upper $3\times 3$ block of the winter season TPM. Green is the extracted value, orange is the Markov state model. \textbf{Bottom}: Upper $3\times 3$ block of the memory kernel for the above TPM, showing the delta-like component and then noise around zero. Choosing $\tau=1$ gives the orange MSM in the top panel.}
    \label{fig:winter_TPM}
\end{figure}

For the seasonal state models, the TPMs and memory kernels were noisy. We show the upper 3x3 block of the TPM for the winter season in Fig.~\ref{fig:winter_TPM}-top. In each season, $\mc{K}_{ij}(1)\gg\mc{K}_{ij}(t>1)$, implying that the Markovian component dominates the memory. We show this for the winter season in Fig.~\ref{fig:winter_TPM}-bottom. Error plots confirmed that the Markovian approximation gives an error that is either better or equivalent to including more data. 
Thus, we took the Markovian limit of $\tau=1$, which resulted in the predicted TPM elements shown in Fig.~\ref{fig:winter_TPM}-top.
\subsection{Bistable Potential Simulation}\label{app:double_well}
We generate Langevin dynamics using a Python script,
\vspace{-4pt}\begin{spverbatim}
rando = np.random.randn(n-1)
for i in range(n - 1):
    x[i + 1] = x[i] + v[i] * dt 
    v[i + 1] = v[i] + dt * (
                -1 * sigmoid.f(x[i])**2 * 
                (\sigma_{bis}**2/2/kbT) * v[i] 
                + 4 * x[i] * (1 - x[i] ** 2)
                ) 
                + sigmoid.f(x[i]) * sigma_bis 
                * sqrtdt * rando[i]
\end{spverbatim}\vspace{4pt}
\noindent where we introduce the position-dependent friction by sigmoid.f, a class that rounds the position to the nearest 1E-4 and caches results for lookup. Due to the relation between the friction and the random force, the scaling appears squared in the memory term. The values of the parameters were 
\vspace{-4pt}\begin{spverbatim}
kbT = .5  # beta^-1 in Joules
sigma_bis = np.sqrt(2. * kbT)
dt = .01  # Time step in seconds
sqrtdt = np.sqrt(dt) 
T = 1E6  # Total time in seconds
n = T/dt
\end{spverbatim}\vspace{4pt}

\vfill\onecolumngrid
\pagebreak
\twocolumngrid
\section*{References}
\vspace{-14pt}
\bibliography{export}

\onecolumngrid
\vspace{20pt}
\hrulefill

\section*{Supporting Information}

\section{Generalized Langevin Equations}\label{app:GLE}

The GLE is a general---indeed, formally exact---result from projection operator theory \cite{BookZwanzig}. In the standard, equilibrium derivation, one considers only tracking a small part of a larger, closed space that evolves according to a Markovian equation of motion. This could be carving out a smaller, open system in the style of system-bath Hamiltonians that form the basis of perturbation theory approaches, or it could be a projection that considers the whole spatial extent but selects only a subset of the possible observables and discards the others. The operator $\mc{P}$ is idempotent, $\mc{P}^2 = \mc{P}$, acts in the full space to give the projected quantity, and can be used to define its orthogonal complement, $\mc{Q}\equiv\mathbb{1}-\mc{P}$. Upon acting with the projector, one is required to take an average with respect to a particular distribution, enabling this framework to span both equilibrium and nonequilibrium scenarios.

To derive the GLE, we begin with the Liouvillian definition of the time derivative of some observable, $X$,
\begin{equation}
    \frac{\mathrm{d}}{\mathrm{d}t}X(t)=\e{\imi\mc{L}t}\imi\mc{L}X,
\end{equation}
where $\mathcal{L}$ is the Liouvillian generator of the dynamics: in quantum mechanics, this is the commutator with the Hamiltonian operator; in classical mechanics, this is the Poisson bracket. We seek the equation of motion in terms of contributions from the two regions defined by $\mathcal{P}$ and $\mathcal{Q}$. Inserting the resolution of the identity, $\mathbb{1} = \mathcal{P} + \mathcal{Q}$,
\begin{equation}
     \frac{\mathrm{d}}{\mathrm{d}t}X(t) = \e{\imi\mc{L}t}(\mc{P}+\mc{Q})\imi\mc{L}X,
\end{equation}
and then expanding the propagator in the second term using the Dyson identity,
\begin{align}
    &\frac{\mathrm{d}}{\mathrm{d}t}X(t) = \e{\imi\mc{L}t}\mc{P\imi L}X \nonumber\\
    &\ \ - \int_0^t\dd{s}\e{\imi\mc{L}s}\mc{PL}\e{\imi\mc{QL}(t-s)}\mc{QL}X  + \e{\imi\mc{QL}t}\mc{Q\imi L}X, \tag{\ref{eq:GLE}}
\end{align}
we recover Eq.~\ref{eq:GLE_approx} in the main text. Different projectors generate distinct instantiations of the GLE. 

Below, we illustrate how freedom in the projection operator allows us to obtain various versions of the exact GLE (Mori, Zwanzig, etc.) for a generalized velocity-like observable and review commonly adopted simplifications that render the GLE simpler to parameterize. We delineate the derivation and, when widely adopted, approximations used in various GLEs that describe the evolution of the velocity of a tagged particle in a fluid, i.e., we select  $X \equiv \dot{x}$. This example enables us to connect to the wider literature in chemical physics and fluid dynamics. It also provides a suggestive form that serves as the starting point for considering the GLE obeyed by the temperature fluctuations and their derivative, $T$ and $\dot{T}$. 

\subsection{Exact and approximate GLEs for a tagged particle}\label{GLE-forms}

\textbf{Mori GLE}. The Mori projector generally spans an observable, $B$, and its time derivative, $\dot{B}$, with the inner product generally defined over the equilibrium distribution (or any other time-translationally invariant distribution) that causes the members of the projector to be orthogonal, and acts on observable $X$:
\begin{equation}
    \mc{P}_M X(t) = \frac{\langle B(0)|X(t)\rangle}{\langle B(0)\rangle^2}B(0) + \frac{\langle \dot{B}(0) | X(t)\rangle}{\langle \dot{B}(0)\rangle ^2}\dot{B}(0).
\end{equation}
Taking $X$ to be the velocity of a tagged particle, $v = \dot{x}$, and $B \equiv x$, we find its Mori GLE,
\begin{equation}\label{eq:GLE_mori}
    \ddot{x}_t = -\Omega (x_t-x_{eq}) - \int_0^t\dd{s}\dot{x}_s\Gamma_\mathrm{M}(t-s) + f_\mathrm{M}^\mathrm{R}(t).
\end{equation}
The first (spring) term is often written as $K={\langle \dot{x}^2\rangle}/{\langle x^2\rangle}$. $f^R(t)$ is the `random' force, which is a deterministic function that depends on the initial distribution of the full system's coordinates, $\chi$, and captures all non-linearities. \textbf{Common approximations:}  $f^R(t)$ is often replaced by a colored, Gaussian noise, which is only appropriate if the system behaves linearly.

\textbf{Zwanzig GLE}. Here, we employ the Zwanzig projector, which performs a conditional average predicated on a particular value of observables $B$ and $\dot{B}$ as functions of the entire phase space variables at $t=0$, $\chi_0$,
\begin{equation}
\begin{split}
    &\lr{X(t)}_{B_0, \dot{B}_0} = \mc{P}_Z X(t) \\
    &\ \ = \frac{ \lr{\delta[B(\hat{\chi}_0) - B(\chi)] \delta[\dot{B}(\hat{\chi}_0) - \dot{B}(\chi)] , X(\hat{\chi}_0, t) } }{ \lr{ \delta[B(\hat{\chi}_0) - B(\chi)] \delta[\dot{B}(\hat{\chi}_0) - \dot{B}(\chi)] }}.
\end{split}
\end{equation}
Here, the inner product requires performing an integral over $\chi_0$ and an average over a distribution $f(\chi_0)$, with the hat distinguishing the integrated variable. With this projector, the Zwanzig GLE for $\dot{x}_t$ takes the form,
\begin{equation}\label{eq:GLE_zwanzig_mass}
\begin{split}
       \ddot{x}_t &= -\frac{1}{M(x_t)}\frac{\mathrm{d}}{\mathrm{d}x_t}\left[ \Phi_\mathrm{MF}(x_t) + \log(M(x_t)) \right] \\
       &\hspace{5pt}- \int_0^t\dd{s}\left[\frac{\nabla_{\dot{x}}}{\beta M(x_s)}-\dot{x}_s\right]\Gamma_\mathrm{Z}(t-s, x_s, \dot{x}_s) + f_\mathrm{Z}^\mathrm{R}(t) , 
\end{split}
\end{equation}
where the PMF, $\Phi_\mathrm{MF}(x_t)$, is the histogram $-\log\mathbb{P}(x_t)$, and the term in square brackets is the `effective' potential, $\Phi(x_t)$. We note that the Zwanzig friction kernel generally depends on the position and momenta. \textbf{Common approximations:}
The mass is often independent of position. Indeed one can analytically show it is a constant for phase-space coordinates and certain nonlinear observables.\cite{Vroylandt2022b, Ayaz2022} In this case the effective potential resolves to the potential of mean force,
\begin{equation}\label{eq:GLE_zwanzig}
\begin{split}
       \ddot{x}_t &= -\frac{\nabla_x \Phi_\mathrm{MF}(x_t)}{M} \\
       &\hspace{5pt}- \int_0^t\dd{s}\left[\frac{\nabla_{\dot{x}}}{\beta M}-\dot{x}_s\right]\Gamma_\mathrm{Z}(t-s, x_s, \dot{x}_s) + f_\mathrm{Z}^\mathrm{R}(t).
\end{split}
\end{equation}
Further, the friction kernel may also be taken as independent of position \textit{and} momentum, although no similar analytical justification holds. Neglecting these dependencies, Eq.~\ref{eq:GLE_zwanzig} reduces to the well-known and commonly adopted form,
\begin{equation}
    \ddot{x}_t = -\frac{\nabla_x \Phi_\mathrm{MF}(x_t)}{M} - \int_0^t\dd{s}\dot{x}_s\Gamma_\mathrm{app}(t-s) + f_\mathrm{Z}^\mathrm{R}(t) ,    \tag{\ref{eq:GLE_approx}}
\end{equation}
which is no longer exact and hence referred to as the `approximate' GLE.

\textbf{Minimal GLE.} One can go beyond these projectors and consider an entire manifold of choices for the basis,\cite{Vroylandt2022b} including a combination\cite{Ayaz2022} of the previous two. From a data-driven perspective, the simplest choice is to use the PMF itself as the basis function, upon which the ``minimal'' reads as
\begin{align}\label{eq:GLE_vroylandt}
    \ddot{x}_t &= -\frac{\nabla_x \Phi_\mathrm{MF}(x_t)}{M} \nonumber\\
    &\hspace{5pt} - \int_0^t\dd{s}\dot{x}_s\nabla^2\Phi_{\mathrm{MF}}(x_s)\Gamma_\mathrm{min}(t-s)  + f_\mathrm{min}^\mathrm{R}(t).
\end{align}
That is, one assumes that the position dependence of the memory kernel is separable, leading to a term that goes as the curvature of the PMF. For a full derivation, see Ref.~\onlinecite{Vroylandt2022b}. \textbf{Common approximations:} The minimal GLE is derived under the assumption of constant mass. 

\subsection{Apparent Potential}\label{app:GLE-app-pot}

In general, the mass is not a constant with position, but one may decide to invoke this position independence of the mass as an approximation. In this case, the first term on the right-hand side of the GLE expression---the effective force term (Eq.~\ref{eq:GLE_zwanzig_mass})---would be reduced to the mass independent expression (Eq.~\ref{eq:GLE_zwanzig}). That is 
\begin{equation}
    -\frac{\nabla_x\Phi_\mathrm{app}(x_t)}{M}\equiv -\frac{\nabla_x\Phi(x_t)}{M(x_t)},
\end{equation}
where $M$ is the value taken for the constant mass (probably from equipartition) and due to the approximation, $\Phi_\mathrm{app}$ is not measured PMF. Instead, one might derive the force or use some model function, e.g., the Mori (harmonic) form. If one uses the PMF directly, it would lead to some error. To quantify the error incurred in this case, one must appreciate that the potential of the effective force $\Phi(x)$ is not the \textit{only} source of position dependence, since one must also consider the mass denominator. In other words
\begin{equation}
    \Phi_\mathrm{app}(x) = M\int \dd{x'} \frac{\nabla_{x'}\Phi(x'_t)}{M(x'_t)}
\end{equation}
is the apparent potential whose derivative gives the conservative force. This potential is compared to the PMF in Fig.~\ref{fig:GLE}. If $\Phi_{\rm app}$ is more Gaussian than the $\Phi_{\rm MF}$, then the combined approximations of constant mass and harmonic potential enjoy a serendipitous cancellation of errors that moves the system closer to the Mori GLE in Eq.~\ref{eq:GLE_mori}. This alignment appears to be accidental, but may also hint at the applicability if the Mori GLE, even when it is not immediately obvious. It bears emphasizing that the Mori GLE in Eq.~\ref{eq:GLE_mori} arises from the choice of projector, is exact, and does not require the aforementioned approximations. However, the resulting random force can be a complicated object, and the next step in the usual workflow, which is to replace it with normally distributed random numbers, is not always valid.

\section{2D State Aggregation}

\begin{figure*}[!ht]
    \centering
    \hspace{-15pt}
    \begin{subfigure}[b]{0.65\textwidth}
        \resizebox{\textwidth}{!}{\includegraphics{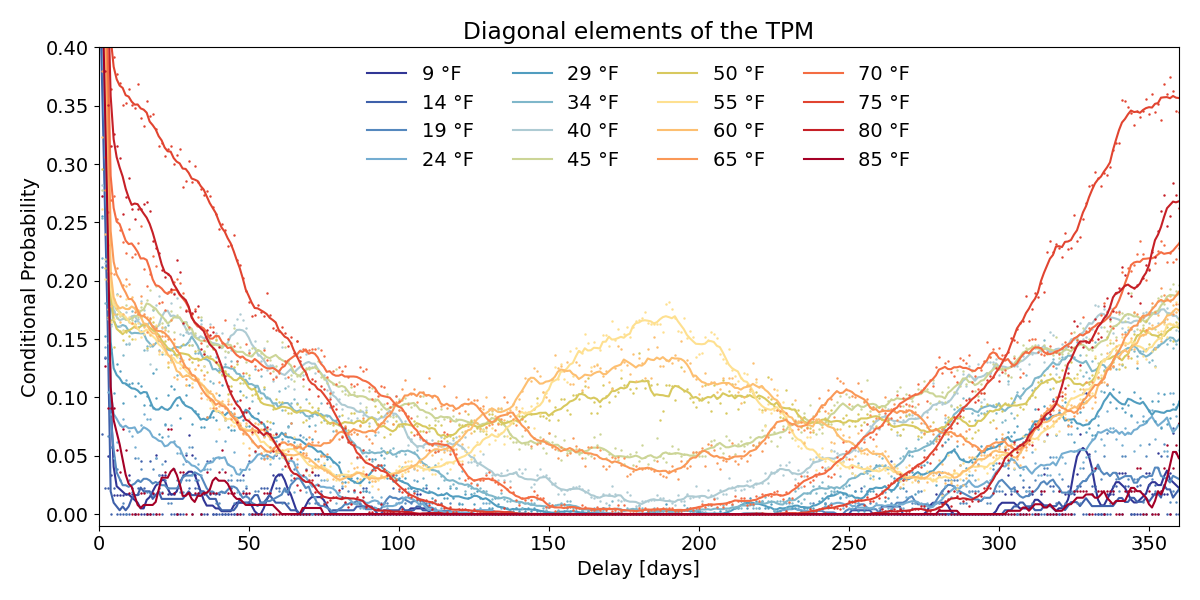}} 
        \caption{}
        \label{fig:FTPM}
    \end{subfigure}
    \\
    \vspace{-5pt}
    \begin{subfigure}[b]{0.4\textwidth}
        \resizebox{\textwidth}{!}{\includegraphics{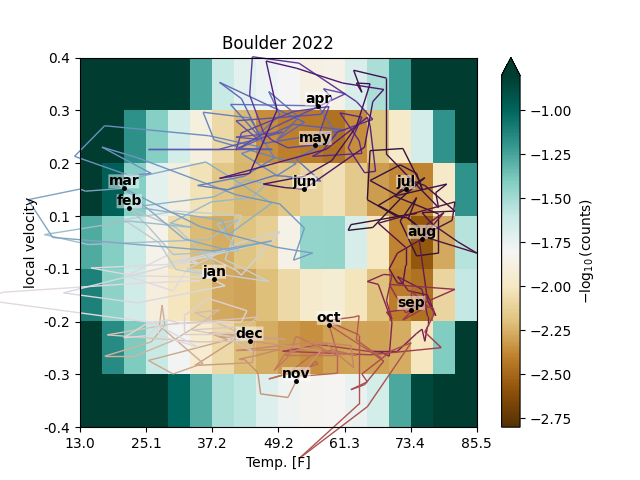}} 
        \caption{}
        \label{fig:donut}
    \end{subfigure}
    \hspace{-5pt}
    \begin{subfigure}[b]{0.425\textwidth}
        \resizebox{\textwidth}{!}{\includegraphics{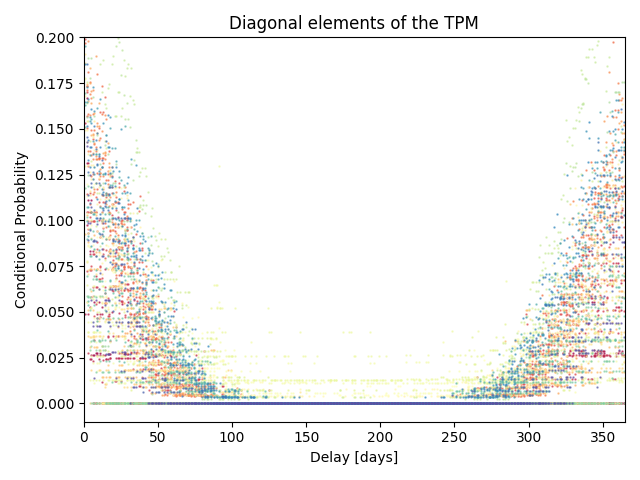}} 
        \caption{}
        \label{fig:gap}
    \end{subfigure}
    \\    
    \begin{subfigure}[b]{0.45\textwidth}  
        \resizebox{\textwidth}{!}{\includegraphics{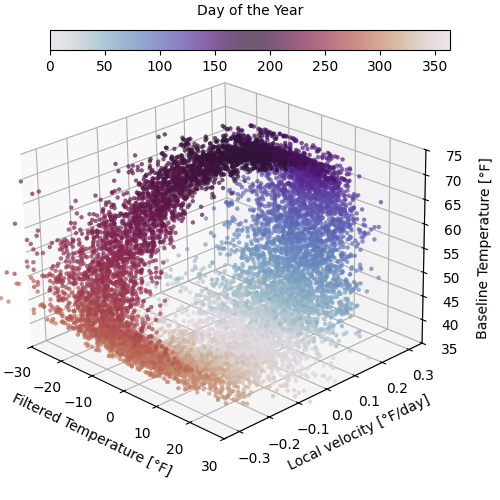}} 
        \caption{}
        \label{fig:bagel}
    \end{subfigure}
    \vspace{-5pt}
    \caption{Attempts to move weather time series into a state picture. a) Partitioning the temperature series into regular intervals. The periodicity is fatal because the half-year recurrence comes in before the summer/winter temperatures decorrelate. b) 2D projection taking also the derivative of the temperature locally, creating a donut with 3 basins. c) Diagonal elements of the TPM from the 2D partitioning in panel (b) are more dissipative than panel (a), but noisier, and still have non-vanishing elements representing the steady-state contribution. d) Separation into a 2D baseline plus fluctuations (filtered). It can be seen that the range of temperatures is compressed in summer (purple/top) compared to very large deviations in winter (white/bottom). To be compared with the predicted version Fig.~\ref{fig:vegan_bagel}}
    \label{fig:2d_states}
\end{figure*}

One could consider a hybrid approach: augmenting the data series to include a measure of the position in the year (a local velocity). In this 2D approach, we can separate summer from winter, unlike in a filtering approach. Unfortunately, dividing the data into higher dimensions has seriously deleterious scaling for measuring transitions and quickly results in unconverged statistics. This could be compensated by a gain in efficiency since the new TPM decays much faster compared to the 1D case: the mixing of autumn and spring in the 1D case leads to longer memories since their temperature distributions evolve in opposite directions but are treated as equivalent in a 1D memory. However, describing autumn and spring separately only partially solves the problem, as we still cannot properly distinguish fluctuations from the periodic, steady-state motion between extremes within one season (see Fig.~\ref{fig:donut}). That is, does a hot day in spring mean a fluctuation, or is it the onset of summer? This is exactly why the filtering was necessary in the first place. 


\vfill

\vfill
\pagebreak

\vfill

\clearpage
\clearpage

\end{document}